\documentclass[a4paper,11pt]{article}

\usepackage{graphicx}% Include figure files
\usepackage{caption}
\usepackage{subcaption}
\usepackage{dcolumn}% Align table columns on decimal point
\usepackage{bm}% bold math
\usepackage{float} % 
\usepackage[export]{adjustbox}
\usepackage{amsmath}
\usepackage{hyphenat}

\pdfoutput=1 % if your are submitting a pdflatex (i.e. if you have
\usepackage{jcappub} % for details on the use of the package, please
\usepackage[T1]{fontenc} % if needed

\title{Loss-Cone–Limited Dark Matter Capture by Early Black Hole Seeds}

\author{Brian Zhang}
\author{and Grant J. Mathews}
\affiliation{Center for Astrophysics, Department of Physics and Astronomy, University of Notre Dame,\\Notre Dame, Indiana, 46556, USA}

\emailAdd{bzhang8@nd.edu}
\emailAdd{gmathews@nd.edu}

\abstract{The rapid appearance of super-massive black holes at high redshift motivates a reassessment of non-baryonic growth channels. We develop a new loss-cone framework for collisionless cold dark-matter (CDM) capture by early black-hole seeds, with particular attention to phase-space depletion and refilling. We model the DM halo with an isotropic distribution function obtained from an NFW-like density profile via Eddington inversion.  We impose direct capture through a relativistic angular-momentum boundary, \(J_{\rm lc}\simeq 4GM_{\rm BH}/c\). We compute the capture rate with an orbit-averaged 1D
Fokker–Planck equation. Refilling by massive perturbers, especially primordial black holes (PBHs), is controlled by the granularity parameter \(\Xi\equiv f_{\rm PBH}m_{\rm PBH}\), while collisionless refilling by triaxial/chaotic centrophilic orbits is bracketed phenomenologically through a parameter \(f_{\rm ch}\). 
We show that ordinary stellar relaxation gives negligible DM-driven growth for our fiducial high-redshift BH seeds, whereas PBH-driven granularity can yield an order-of-magnitude increase growth rate in for sufficiently compact halos. Triaxial or full-loss-cone supply can produce rapid early growth, but the self-consistent evolution generally becomes supply-limited. Once the accessible low-angular-momentum phase-space reservoir is depleted, the capture rate collapses and the black-hole mass saturates. Comparing fixed-density NFW calculations with self-consistent distribution-function evolution, we find that fixed-background models overestimate sustained growth, especially in the full-loss-cone limit. As a cosmological benchmark, we apply the same model to a TNG50-calibrated NFW profile at \(z=20\) and find negligible growth even under optimistic refilling assumptions. Thus, collisionless DM capture is unlikely to solve early SMBH growth in generic NFW-like halos.  However,  it can provide a rapid, non-radiative upper-envelope contribution in rare compact environments with efficient angular-momentum refilling.}


\begin{document}
\maketitle
\flushbottom

\section{Introduction} 

The discovery of accreting black holes (BHs) at very high redshift has exacerbated the
early supermassive\hyp{}black\hyp{}hole (SMBH) growth problem.  The {\it James Webb Space Telescope} (JWST) spectroscopy and
multi-wavelength follow-up have revealed low-mass and overmassive active black
holes at \(z\gtrsim 5\), including systems whose inferred black-hole masses are
large compared with their host stellar masses \cite{kocevski2023,Greene2023,
Natarajan24,Goulding2023,Maiolino_2024,Juod_balis_2024,scholtz2025}.  These objects do not
by themselves select a unique formation channel, but they reduce the time
available for the formation of seed BHs and subsequent growth.  They therefore motivate
models in which massive BH seeds form early, or in which alternative  growth channels
operate before ordinary luminous accretion becomes efficient
\cite{Inayoshi2020,Volonteri2021,regan2024,Kritos2024,kritos2024supe,Imai2026}.

Several channels have been proposed for the formation of BH seeds.  Light seeds may arise as remnants of
metal-poor Population III stars \cite{Xu_2013,Susa_2014,klessen2023}, while intermediate
or massive BH seeds can form through runaway stellar collisions in dense clusters
\cite{sakurai2017,reinoso2018,Portegies02,Kritos2024,kritos2024supe,liempi2024,gonzalez2024,vergara2026}.  Heavy seeds may also
form through direct collapse in metal-poor atomic-cooling halos, especially
when fragmentation is suppressed \cite{BrommLoeb2003,Begelman2006,Agarwal_2014,
Latif2022}.  In cosmological simulations, seed black holes are commonly inserted by hand
into sufficiently massive halos because there is no adequately resolved  sub-grid representation of the SMBH
formation physics \cite{Weinberger17,Pillepich:2019,Nelson:2019}.  The central
uncertainty is not only how the first seed forms, but also how much mass it can
gain in the short interval between \(z\sim 30\) and the observed \(z\gtrsim 7\)
population.

Cold Dark matter (CDM), possibly including primordial black holes (PBHs),  provide a possible
non-baryonic contribution to this early growth. An equilibrium collisionless CDM halo has no steady radial inflow, and sustained black-hole growth occurs only through capture of CDM scattered or torqued into the relativistic loss cone. Although CDM does not accrete like normal matter in the sense of collapse by radiative cooling, it can gravitationally capture onto a BH seed if its angular momentum is low enough. This gravitational capture and its effect on the CDM distribution function is the subject of the current manuscript.   

PBHs are a long-standing
candidate for part of the DM abundance \cite{Hawking1971,Carr2016,Carr_2021,bird2023,Correa2022},
and they have also been discussed as possible seeds for SMBH formation
\cite{ziparo2024,dayal2024}.  Even when PBHs make up only a subdominant fraction of the
CDM, a young black-hole seed is expected to reside inside a collisionless DM
halo that may also contain a granular compact-object component. Here, the term {\it granular}, refers to PBhs, stars, remnants, etc. that may themselves also capture DM halos. In the present work we show the new result that this granularity may significantly increase the SMBH growth rate.  

Gravitational capture of this
ambient material would not radiate and could therefore supplement, or even 
precede, gas accretion because there is no cooling time scale. Previous estimates of DM-assisted SMBH growth have
therefore explored whether dense central DM configurations, possibly with a  PBH component,
can accelerate early black-hole assembly \cite{Gondolo1999,Ferrer2017,
BertoneMerritt2005,Shapiro2023,Imai2026}. A key issue, however, is how rapid depletion of the low angular momentum component of the CDM affects its distribution function.  This has not been considered in previous treatments of DM capture and a new feature of the present work is to explicitly address this issue.

The key dynamical question is not simply whether DM or PBHs pass near the BH seed 
but rather whether the low-angular-momentum phase space is continuously supplied to the
capture boundary.  In stellar dynamics this is the classical loss cone problem:
orbits inside the loss cone are rapidly removed, while two-body relaxation,
massive perturbers, or non-spherical torques determine how quickly they are
replenished \cite{FrankRees1976,LightmanShapiro1977,CohnKulsrud1978,
merritt2005,Perets2007,Vasiliev2014}. Gravitational capture of collisionless DM by a black hole
has the same supply limitation.  A fixed ambient density or a purely geometric
capture cross section can overestimate long-term growth if the accessible
low-angular-momentum reservoir is depleted faster than it is refilled.  This
point is especially important for compact high-redshift halos, where an
optimistic instantaneous capture rate can coexist with a finite phase-space
reservoir. Therefore, in this paper we develop a new loss-cone application  for a solution to the collisionless diffusion equation describing the evolution of the distribution function for CDM due to capture by
early black-hole seeds. 

We initialize the halo with an isotropic distribution
function obtained from a truncated NFW-like density profile through Eddington
inversion, following the standard construction of self-consistent spherical
halo models \cite{nfw,Widrow2000,Kazantzidis2004}.  Direct gravitational capture is imposed
through a relativistic angular-momentum boundary,
\(J_{\rm lc}\simeq 4GM_{\rm BH}/c\), and the loss-cone flux is computed with an
orbit-averaged one-dimensional Fokker--Planck approximation to the Boltzmann equation.  We include ordinary
stellar relaxation as a baseline.  We also consider  PBHs as massive perturbers that can refill the CDM
loss cone, and we consider a phenomenological collisionless refilling channel associated
with triaxial or chaotic centrophilic orbits
\cite{merritt2004,merritt2011,gualandris2017,Vasiliev2014}.  The PBH-driven
diffusion strength is summarized by the granularity parameter
\(\Xi\equiv f_{\rm PBH}m_{\rm PBH}\), while the uncertain non-spherical supply is
bracketed by a centrophilic fraction \(f_{\rm ch}\).

Our main aim and the new result in this work is to determine when this dark growth channel is supply-limited.
We therefore compare fixed-background calculations with a self-consistent treatment of the 
evolution of the halo distribution function and potential.  The new resulting
growth histories show that ordinary stellar relaxation gives negligible
DM-driven growth for the fiducial high-redshift BH seeds.  However, PBH-driven granularity
can produce an order-of-magnitude increase in the growth rate for  sufficiently compact halos, and
triaxial or full-loss-cone supply can trigger a rapid early burst.  However, the
self-consistent models generically saturate once the accessible phase-space
reservoir is depleted.  

As a cosmological benchmark, we apply the same machinery
to a TNG50-calibrated NFW profile at \(z=20\) and find negligible growth even
under optimistic refilling assumptions.  Thus, collisionless CDM gravitational capture is
unlikely to solve early SMBH assembly in generic NFW-like halos, but it provides
a useful upper-envelope contribution in rare compact environments with efficient
angular-momentum refilling (cf. \cite{Imai2026}).

\section{Physical Setup and Phase-Space Description}
\label{sec:setup}

\subsection{Fiducial seed scenario and host halo}
\label{subsec:seed_halo}

We consider the growth of a black-hole seed of mass $M_{\mathrm{BH}}(t)$ embedded in a spherically
symmetric dark-matter (DM) halo at high redshift. In this context, DM denotes the combined population
of particle DM and PBHs. Throughout this manuscript, an overdot denotes a time derivative, e.g.
$\dot{M}_{\mathrm{BH}} \equiv dM_{\mathrm{BH}}/dt$.

The total gravitational potential is written as
\begin{equation}
\Phi(r,t) = \Phi_{\rm h}(r,t) - \frac{G\,M_{\mathrm{BH}}(t)}{r},
\label{eq:Phi_total}
\end{equation}
where $\Phi_{\rm h}$ is the halo potential sourced by the (remaining) DM distribution.
To define a positive binding energy, we use the standard \emph{relative} potential
\begin{equation}
\psi(r,t) \equiv -\Phi(r,t) + \Phi(r_{\max},t),
\qquad \psi(r_{\max},t) = 0,
\label{eq:psi_def}
\end{equation}
where $r_{\max}$ is an outer truncation radius chosen large enough that $\rho(r_{\max})$ is negligible.

\subsection{Isotropic distribution function and Eddington inversion}
\label{subsec:eddington}

We model the DM as a collisionless phase-space distribution function (DF) that is isotropic in
velocity space outside a narrow loss-cone boundary layer. The DF therefore depends on the relative
energy per unit mass,
\begin{equation}
\epsilon \equiv \psi(r,t) - \frac{v^2}{2} \ge 0,
\label{eq:epsilon_def}
\end{equation}
so that $f = f(\epsilon,t)$. For an isotropic DF in a spherical potential, the mass density is the
velocity integral
\begin{equation}
\rho(r,t) = \int f(\epsilon,t)\,d^3v
= 4\sqrt{2}\,\pi \int_{0}^{\psi(r,t)} f(\epsilon,t)\,\sqrt{\psi(r,t)-\epsilon}\; d\epsilon.
\label{eq:rho_from_f}
\end{equation}

We take a truncated NFW halo~\cite{nfw} as the initial density profile to ensure finite mass and a well-behaved DF,
\begin{equation}
\rho(r,t_i)=\rho_{\rm NFW}(r)\,T(r),\qquad
\rho_{\rm NFW}(r)=\frac{\rho_s}{(r/r_s)(1+r/r_s)^2},
\label{eq:truncated_NFW}
\end{equation}
e.g. with a smooth truncation $T(r)=\exp[-(r/r_t)^2]$ and $r_t\sim r_{\rm vir}$. The corresponding isotropic DF may be
obtained via Eddington inversion~\cite{sanchez2023},
\begin{equation}
f(\epsilon,t_i)=\frac{1}{\sqrt{8}\,\pi^2}
\left[
\int_{0}^{\epsilon}\frac{d^2\rho/d\psi^2}{\sqrt{\epsilon-\psi}}\;d\psi
+\frac{1}{\sqrt{\epsilon}}
\left(\frac{d\rho}{d\psi}\right)_{\psi=0}
\right]_{t=t_i},
\label{eq:eddington}
\end{equation}
where $\rho(\psi)$ is understood as the initial density profile expressed as a function of the
initial relative potential. For a sufficiently smooth outer truncation the boundary term is negligible.

A central role in what follows is played by the \emph{phase-space density of states} (phase volume per
unit energy),
\begin{equation}
g(\epsilon,t)
\equiv \int d^3x\,d^3v\;
\delta\!\left(\epsilon-\psi(r,t)+\frac{v^2}{2}\right)
= 16\pi^2\int_{0}^{r_{\max}(\epsilon,t)} r^2\sqrt{2\,[\psi(r,t)-\epsilon]}\;dr,
\label{eq:g_def}
\end{equation}
where $r_{\max}(\epsilon,t)$ is defined implicitly by $\psi(r_{\max},t)=\epsilon$.
The DM mass per unit energy is then
\begin{equation}
\frac{dM(\epsilon,t)}{d\epsilon} = g(\epsilon,t)\,f(\epsilon,t).
\label{eq:dMde}
\end{equation}

\subsection{Angular momentum, circular-orbit scale, and the loss cone}
\label{subsec:losscone_setup}

At fixed $\epsilon$, bound orbits are labeled by their specific angular momentum $J$, with
$0\le J \le J_c(\epsilon,t)$. Here $J_c(\epsilon,t)$ is the maximum angular momentum allowed for bound
motion at energy $\epsilon$, i.e., the circular orbit. We define a dimensionless angular-momentum coordinate
\begin{equation}
R \equiv \frac{J^2}{J_c^2(\epsilon,t)} \in [0,1].
\label{eq:R_def}
\end{equation}

Collisionless capture by the central BH is implemented as an absorbing boundary in $J$ (or $R$).
For a Schwarzschild BH, a GR-motivated angular momentum threshold for direct capture of a marginally bound
(\emph{effectively parabolic}) orbit is~\cite{shapiro2024}
\begin{equation}
J_{\rm lc} \simeq \frac{4GM_\mathrm{BH}}{c},
\label{eq:Jlc}
\end{equation}
which defines the \emph{loss cone} $J<J_{\rm lc}$. In terms of $R$,
\begin{equation}
R_{\rm lc}(\epsilon,t) \equiv \frac{J_{\rm lc}^2}{J_c^2(\epsilon,t)}.
\label{eq:Rlc_def}
\end{equation}

 Note that Eq.~(\ref{eq:Jlc}) can be related to a cross section for gravitational capture \cite{shapiro2024} of a CDM particle as shown below.

In the BH-dominated (Keplerian) region, $\psi(r)\simeq GM_\mathrm{BH}/r$ and the circular-orbit relations
simplify to
\begin{equation}
J_c(\epsilon) = \frac{GM_\mathrm{BH}}{\sqrt{2\epsilon}},
\qquad
P(\epsilon) = \frac{\pi GM_\mathrm{BH}}{\sqrt{2}\,\epsilon^{3/2}},
\qquad
R_{\rm lc}(\epsilon)=32\,\frac{\epsilon}{c^2}\ll 1,
\label{eq:kepler_relations}
\end{equation}
where $P(\epsilon)\equiv\oint dr/|v_r|$ is the (radial/orbital) period of the particle with radial velocity
$v_r$. The density of energy states follows from Eq.~\eqref{eq:g_def},
\begin{equation}
    g(\epsilon,t)=4\pi^2 P(\epsilon)J_c^2.
    \label{eq:g_simplified}
\end{equation}

\section{Loss-Cone Refilling and the DM Capture Rate}
\label{sec:losscap}
\subsection{Angular-momentum diffusion}
\label{subsec:Angular-momentum_diffusion}

Many previous studies have examined the growth of SMBHs resulting from the collisionless capture of DM. However, collisionless capture is fundamentally phase-space limited: a black hole can capture particles already inside its loss cone, but after those orbits are depleted, sustained capture requires angular-momentum refilling by gravitational encounters or non-spherical torques.
Loss-cone refilling can be treated as a cumulative transport process produced by many discrete weak
gravitational encounters induced by a collisionless population of massive perturbers (e.g. stars, stellar remnants, or PBHs).  Loss-cone refilling is therefore not
described by a single refilling cross section.  Instead, the cross section
for an individual gravitational encounter determines velocity-space
diffusion coefficients, which in turn determine diffusion in angular
momentum \citep{CohnKulsrud1978,merritt2013}.

Consider a particle-DM test particle of mass $m_\chi$ moving relative to a perturber of mass $m_p$ with asymptotic relative speed
\begin{equation}
u \equiv |\boldsymbol{v}_\chi-\boldsymbol{v}_p|.
\end{equation}
For an isolated Newtonian hyperbolic encounter with impact parameter $b$,
the deflection angle $\Theta$ obeys
\begin{equation}
\tan\left(\frac{\Theta}{2}\right)
=
\frac{G(m_p+m_\chi)}{b u^2}.
\label{eq:b_theta_grav}
\end{equation}
The corresponding gravitational Rutherford differential cross section is
\begin{equation}
\frac{d\sigma_{\rm grav}}{d\Omega}
=
\left[
\frac{G(m_p+m_\chi)}{2u^2}
\right]^2
\frac{1}{\sin^4(\Theta/2)}.
\label{eq:grav_rutherford}
\end{equation}
For particle DM, $m_\chi\ll m_p$, and in the small-angle limit
$b\gg Gm_p/u^2$, Equations~\eqref{eq:b_theta_grav} and
\eqref{eq:grav_rutherford} imply
\begin{equation}
\Theta \simeq \frac{2Gm_p}{b u^2},
\qquad
|\Delta \boldsymbol{v}_{\perp}|
\simeq \frac{2Gm_p}{b u},
\label{eq:small_angle_kick}
\end{equation}
where $\Delta \boldsymbol{v}_{\perp}$ is the velocity kick perpendicular to
the unperturbed relative motion.

Let the perturbers have number density $n_p=\rho_p/m_p$.  Encounters with
impact parameters between $b$ and $b+db$ occur at the rate
\begin{equation}
d\Gamma = 2\pi b\,db\,n_p u.
\end{equation}
Consequently, the transverse velocity-diffusion rate in the mono-velocity
approximation is
\begin{align}
\mathcal{B}_{\perp}
&\equiv
\lim_{\Delta t\rightarrow 0}
\frac{\left\langle
(\Delta\boldsymbol{v}_{\perp})^2
\right\rangle}{\Delta t}
\nonumber\\
&\simeq
\int_{b_{\min}}^{b_{\max}}
(\Delta\boldsymbol{v}_{\perp})^2\,d\Gamma
\nonumber\\
&\simeq
\frac{8\pi G^2m_p^2n_p}{u}
\int_{b_{\min}}^{b_{\max}}\frac{db}{b}
=
\frac{8\pi G^2m_p\rho_p}{u}\ln\Lambda,
\label{eq:velocity_diffusion_scaling}
\end{align}
where
\begin{equation}
\ln\Lambda \equiv \ln\!\left(\frac{b_{\max}}{b_{\min}}\right)
\label{eq:coulomb_log}
\end{equation}
is the Coulomb logarithm, with
\begin{equation}
b_{\min}\sim \frac{G m_p}{\sigma^2(r)},
\qquad
b_{\max}\sim r \ \ \text{(or a characteristic scale of the system)}.
\label{eq:b_min-b_max}
\end{equation}
Since relaxation is achieved when $(\Delta\boldsymbol{v}_{\perp})^2\sim u^2$, this leads to the
standard local (non-resonant) two-body relaxation time for an equal-mass perturber
population with 1D velocity dispersion $\sigma(r)$
\begin{equation}
t_{\rm rel}(r)
\simeq
\frac{0.34\,\sigma^3(r)}
{G^2m_p\rho_p(r)\ln\Lambda(r)},
\label{eq:trel_from_scattering}
\end{equation}
where the numerical coefficient is the standard Maxwellian convention~\cite{spitzer2014,merritt2013}.

During a local encounter, the position is effectively fixed and
\begin{equation}
\Delta\boldsymbol{J}
=
\boldsymbol{r}\times\Delta\boldsymbol{v}.
\end{equation}
In the small-$R$ limit, the standard local diffusion coefficients satisfy
\begin{align}
\mathcal{A}^{\rm loc}_{R}
&\equiv
\lim_{\Delta t\rightarrow0}
\frac{\langle\Delta R\rangle}{\Delta t}
\simeq
\frac{r^2}{J_c^2(\epsilon)}
\mathcal{B}_{\perp},
\label{eq:local_R_drift}\\
\mathcal{B}^{\rm loc}_{RR}
&\equiv
\lim_{\Delta t\rightarrow0}
\frac{\langle(\Delta R)^2\rangle}{\Delta t}
\simeq
2R\,
\frac{r^2}{J_c^2(\epsilon)}
\mathcal{B}_{\perp}.
\label{eq:local_R_diffusion}
\end{align}
These relations show explicitly that the second-order diffusion coefficient
vanishes linearly as $R\rightarrow0$, while
$\mathcal{B}_{RR}/(2R)$ approaches a finite value.

The local coefficient must be averaged over the orbit because the density,
velocity dispersion, and encounter rate vary with radius.  For an orbit
with turning points $r_p$ and $r_a$, radial period
$P_r(\epsilon,J)$, and radial speed $v_r$, the orbit average of any local rate $X(r)$ is
\begin{equation}
\overline{X}(\epsilon,R)
=
\frac{2}{P_r(\epsilon,J)}
\int_{r_p}^{r_a}
\frac{
X(r)
}{
|v_r(r;\epsilon,J)|
}\,dr .
\label{eq:orbit_average}
\end{equation}
Therefore,
\begin{equation}
\overline{\mathcal{B}}_{RR}
=
\frac{2}{P_r}
\int_{r_p}^{r_a}
\frac{
\mathcal{B}^{\rm loc}_{RR}
}{
|v_r|
}\,dr 
\simeq 2R \mu(\epsilon)
,
\label{eq:orbit_average_BRR}
\end{equation}
where
\begin{align}
\mu(\epsilon)
&\equiv
\lim_{R\rightarrow0}
\frac{\overline{\mathcal{B}}_{RR}(\epsilon,R)}{2R}
\nonumber\\
&=
\frac{2}{P_r(\epsilon)J_c^2(\epsilon)}
\int_{r_p}^{r_a}
\frac{r^2\,\mathcal{B}_{\perp}(r)}
{|v_r(r;\epsilon)|}\,dr .
\label{eq:D_orbit_exact}
\end{align}
At fixed energy $\epsilon$, the orbit-averaged
evolution in $R$ can be written as a one-dimensional Fokker–Planck equation in $R$-space,
\begin{equation}
\frac{\partial f(\epsilon,R,t)}{\partial t}
= -\frac{\partial \mathcal{F}_R(\epsilon,R,t)}{\partial R},
\qquad
\mathcal{F}_R(\epsilon,R,t) = -D_{RR}(\epsilon,R)\,\frac{\partial f}{\partial R},
\label{eq:FP_R}
\end{equation}
where $\mathcal{F}_R$ is the flux in $R$, and
\begin{equation}
    D_{RR}=\frac{\overline{\mathcal{B}}_{RR}}{2}
    \label{eq:DRR}
\end{equation}
is the (orbit-averaged) diffusion coefficient. Near the loss cone ($R\ll 1$), Equation~\eqref{eq:orbit_average_BRR} implies 
\begin{equation}
    D_{RR}(\epsilon,R)\simeq\mu(\epsilon)R.
\label{eq:D_RR_linear}
\end{equation}

Since the local velocity-diffusion rate is of order
$\mathcal{B}_{\perp}\sim\sigma^2/t_{\rm rel}$,
Equation~\eqref{eq:D_orbit_exact} implies
\begin{align}
\mu(\epsilon)
&\sim
\frac{2}{P_r(\epsilon)J_c^2(\epsilon)}
\int_{r_p}^{r_a}
\frac{
r^2\sigma^2(r)
}{
|v_r(r;\epsilon)|\,t_{\rm rel}(r)
}\,dr \\
&=
\left\langle \frac{r^2\sigma^2(r)}{J_c^2(\epsilon)}\frac{1}{t_{\rm rel(r)}}\right\rangle_{\rm orb} .
\label{eq:D_from_trel}
\end{align}
Thus $\mu(\epsilon)$ is an orbit-averaged
angular-momentum relaxation rate normalized by $J_c^2(\epsilon)$ and is generally not exactly equal
to the inverse of the local relaxation time.  A useful numerical closure is~\cite{merritt2005}
\begin{equation}
\mu(\epsilon)
\simeq
C_J\,t_{\rm rel}^{-1}
\!\left[r_{\rm char}(\epsilon)\right],
\qquad C_J=\mathcal{O}(1),
\label{eq:D_closure}
\end{equation}
where $r_{\rm char}$ may be taken to be the circular-orbit radius at the same energy (or, alternatively,
the semi-major axis in a Keplerian regime). Where needed, more accurate orbit-averaging can be performed directly
using Eq.~\eqref{eq:D_from_trel}. We adopt $C_J=1$ in the fiducial
calculation and use variations of $C_J$ to characterize the order-unity
uncertainty associated with this closure.

%```

\subsection{Quasi-steady boundary-layer solution and the loss-cone flux}
\label{subsec:quasisteady}

The loss cone typically occupies a very small region of angular-momentum space, $R_{\rm lc}\ll 1$.
Outside of this narrow boundary layer, phase mixing rapidly establishes an approximately isotropic
reservoir at fixed $\epsilon$ with $f(\epsilon,R\simeq 1,t)\equiv f_0(\epsilon,t)$.
If the $R$-profile adjusts rapidly compared to secular changes in $f_0(\epsilon,t)$, the solution in
the boundary layer is quasi-steady, $\partial f/\partial t\simeq 0$, so that $\partial\mathcal{F}_R/\partial R=0$
and $\mathcal{F}_R$ is constant in $R$.

It is convenient to define the \emph{inward} flux magnitude $F(\epsilon,t)>0$ into the loss cone by
\begin{equation}
\mathcal{F}_R(\epsilon,R,t) \equiv -F(\epsilon,t).
\end{equation}
Using Equation~\eqref{eq:D_RR_linear}, Equation~\eqref{eq:FP_R} becomes
\begin{equation}
F = \mu R\,\frac{\partial f}{\partial R}.
\end{equation}
Integrating from $R=R_{\rm lc}$ to $R=1$ yields a logarithmic boundary-layer profile and the relation
\begin{equation}
f_0(\epsilon,t)-f_{\rm lc}(\epsilon,t)
= \frac{F(\epsilon,t)}{\mu(\epsilon)}\,\ln\!\left(\frac{1}{R_{\rm lc}(\epsilon,t)}\right),
\label{eq:flc_relation}
\end{equation}
where $f_{\rm lc}(\epsilon,t)\equiv f(\epsilon,R_{\rm lc},t)$ is the DF value at the loss-cone boundary.

\subsection{Finite draining and the interpolation between full and empty loss cones}
\label{subsec:draining}

Capture is not set solely by diffusion into $R_{\rm lc}$; it also depends on how quickly orbits inside
the loss cone are removed. For collisionless geodesic capture by a BH, an orbit with $R<R_{\rm lc}$
is typically removed on a dynamical time $\sim P(\epsilon)$.
At fixed $\epsilon$, the fraction of angular-momentum phase space inside the loss cone is
$\sim R_{\rm lc}$ because the phase-space measure is uniform in $J^2$ in the Keplerian regime.
The corresponding draining (``pinhole'') flux is therefore
\begin{equation}
F(\epsilon,t) \simeq \frac{R_{\rm lc}(\epsilon,t)}{P(\epsilon,t)}\,f_{\rm lc}(\epsilon,t),
\label{eq:RobinBC}
\end{equation}
which acts as a Robin (radiation) boundary condition connecting the boundary value $f_{\rm lc}$ to
the inward flux~\cite{singer2008,vasiliev2013},
\begin{equation}
    f_{\rm lc}-\mu P\left.\frac{\partial f}{\partial R}\right|_{R_{\rm lc}}\simeq 0.
\label{eq:RobinBCfull}
\end{equation}

Combining Equations~\eqref{eq:flc_relation} and \eqref{eq:RobinBC} gives
\begin{equation}
\frac{f_{\rm lc}}{f_0}
=
\frac{q}{\ln(1/R_{\rm lc})+q},
\qquad
q(\epsilon,t)\equiv \frac{P(\epsilon,t)\,\mu(\epsilon)}{R_{\rm lc}(\epsilon,t)}.
\label{eq:flc_over_f0}
\end{equation}
The capture rate per unit energy is the phase-space flux into the absorbing boundary,
\begin{equation}
\frac{d\dot{M}_\mathrm{BH}(\epsilon,t)}{d\epsilon}
= g(\epsilon,t)\,F(\epsilon,t).
\label{eq:dMdot_gF}
\end{equation}
If the outer isotropic reservoir dominates the halo, then the total DM mass per unit energy is
$dM/d\epsilon \simeq g f_0$. Using Equations~\eqref{eq:RobinBC}--\eqref{eq:dMdot_gF},
the quasi-steady loss-cone capture spectrum becomes:
\begin{equation}
\frac{d\dot{M}_\mathrm{BH}(\epsilon,t)}{d\epsilon}\simeq
\frac{dM(\epsilon,t)}{d\epsilon}\,
\frac{R_{\rm lc}(\epsilon,t)}{P(\epsilon,t)}\,
\frac{q(\epsilon,t)}{\ln\!\left[1/R_{\rm lc}(\epsilon,t)\right]+q(\epsilon,t)}.
\label{eq:dMdot_interp}
\end{equation}

Equation~\eqref{eq:dMdot_interp} smoothly interpolates between:
(i) the \emph{full-loss-cone} (pinhole) limit, $q\gg \ln(1/R_{\rm lc})$, in which the capture is geometric, using Equations~\eqref{eq:dMde}, \eqref{eq:g_simplified}, and \eqref{eq:Rlc_def},
\begin{equation}
\left.\frac{d\dot{M}_\mathrm{BH}}{d\epsilon}\right|_{\rm full}
\simeq
\frac{dM}{d\epsilon}\,\frac{R_{\rm lc}}{P}
\simeq
4 \pi^2 J_{\rm lc}^2 f(\epsilon,t)
,
\label{eq:fullLC}
\end{equation}
and (ii) the \emph{empty-loss-cone} (diffusion-limited) limit, $q\ll 1$, in which the capture rate is
suppressed by the logarithmic barrier,
\begin{equation}
\left.\frac{d\dot{M}_\mathrm{BH}}{d\epsilon}\right|_{\rm empty}
\simeq
\frac{dM}{d\epsilon}\,\frac{\mu(\epsilon)}{\ln(1/R_{\rm lc})}.
\label{eq:emptyLC}
\end{equation}
Note that for an isotropic bath at infinity, $\epsilon=v^2/2$ and $d\epsilon=vdv$. Therefore,
\begin{equation}
    \rho \left\langle\frac{1}{v}\right\rangle=\int f(v,t)\frac{1}{v}d^3v=4\pi\int_0^\infty f(\epsilon,t)d\epsilon.
\end{equation}
The full-loss-cone capture rate is then
\begin{equation}
    \left.\dot{M}_\mathrm{BH}\right|_{\rm full}=4\pi^2 J_{\rm lc}^2\int_0^\infty f(\epsilon,t)d\epsilon=16\pi (G M_{\rm BH})^2 \rho c^{-2} \left\langle v^{-1}\right\rangle,
\end{equation}
 This can be related to a gravitational capture cross section if we write $\dot M_{\rm BH} = \langle \rho \sigma_{\rm capt} v \rangle$.  If $f$ describes a stationary isotropic bath of collisionless  CDM particles with a Maxwell-Boltzmann velocity distribution \cite{shapiro2024}, the gravitational capture cross section is
\begin{equation} \label{9}
\sigma_{\text{capt}} = \frac{8\sqrt{6\pi} R_s^2}{v_{\infty}^2/c^2}~~, 
\end{equation} 
where \(R_s\equiv2GM_{\text{BH}}/c^2\) is the Schwarzschild radius, and 
 $v_{\infty}\ll 1$ is the asymptotic speed of the CDM particle far from the black hole. Alternatively, if $f$ describes a monoenergetic $\delta$-function distribution, then $\sigma_{\rm capt}$ reduces to the geometric capture cross section
$\sigma_{\rm capt}(v)= 4\pi R_s^2 (c/v_{\infty})^2$ \cite{shapiro2024}.

In Section~\ref{sec:results} we evaluate
the total capture rate, $\dot{M}_\mathrm{BH}(t)=\int (d\dot M_\mathrm{BH}/d\epsilon)\,d\epsilon$,
using self-consistent time-dependent halo potentials and several refilling prescriptions for $\mu(\epsilon)$.

\section{Loss-Cone Refilling Mechanisms}
\label{sec:refilling}

In a spherical halo composed solely
of microscopic particle DM, gravitational two-body self-relaxation is
extremely slow, so sustained capture after the initially occupied loss
cone has drained requires an additional source of angular-momentum
transport. Equations~\eqref{eq:dMdot_interp} and \eqref{eq:flc_over_f0} show that the loss-cone refilling depends on an effective
orbit-averaged angular-momentum diffusion rate, $\mu$. In this section we specify $\mu(\epsilon,t)$ for
several physically motivated refilling channels relevant at high redshift. We focus on three
ingredients: (i) stellar two-body relaxation (a baseline collisionless process), (ii) massive
perturbers in a mixed-DM model (PBHs as granularity sources for scattering particle DM), and
(iii) collisionless refilling driven by non-spherical potentials (triaxiality/chaotic
centrophilic orbits).

\subsection{Mixed DM with PBHs as massive perturbers}
\label{subsec:pbh_refill}

We consider a mixed DM model in which the bulk of the mass is in collisionless particle DM, all with mass $m_\chi$, while a separate population of compact-object (PBHs), all with mass $m_{\rm pbh}$, provides granularity (gravitational DM capture) and drives relaxation.
We parametrize the PBH density by
\begin{equation}
\rho_{\rm pbh}(r) \equiv f_{\rm pbh}\,\rho_{\rm DM}(r),
\label{eq:rhoPBH}
\end{equation}
where $\rho_{\rm DM}$ is the total DM density and $0\le f_{\rm pbh}\le 1$ is the fraction of DM in PBHs
with mass $m_{\rm pbh}$.

A key point is that when the test population is \emph{particle DM} and the field population is PBHs, the diffusion rate
is determined by the properties of the \emph{scatterers} (PBHs), not by the test-particle mass. Thus,
in Eq.~\eqref{eq:trel_from_scattering} one should use
\begin{equation}
m_p = m_{\rm pbh},\qquad \rho_p=\rho_{\rm pbh}=f_{\rm pbh}\rho_{\rm DM}.
\label{eq:use_mrho}
\end{equation}
With these substitutions, the PBH-driven relaxation time scales as
\begin{equation}
t_{\rm rel}^{\rm (pbh)}(r)\ \propto\ \frac{\sigma^3(r)}{m_{\rm pbh}\,\rho_{\rm pbh}(r)}
\ \propto\ \frac{\sigma^3(r)}{f_{\rm pbh}\,m_{\rm pbh}\,\rho_{\rm DM}(r)}.
\label{eq:scaling_pbh}
\end{equation}
If $f_{\rm pbh}m_{\rm pbh}\gg(1-f_{\rm pbh})m_\chi$, the presence of massive PBHs can substantially shorten the gravitational relaxation time relative to that of a halo composed entirely of particle
DM. Hence, even a small PBH fraction can dominate the angular-momentum diffusion if $m_{\rm pbh}$ is large
enough, i.e. if there are ``massive perturbers''. Unlike refilling by stars alone, this mechanism does
not require the prior formation of a dense nuclear stellar population
and can therefore operate in star-poor early nuclei. 

The diffusion approximation implicit in Eq.~\eqref{eq:trel_from_scattering} requires that many PBHs contribute
to the stochastic gravitational field within the region that controls loss-cone feeding. A useful
consistency check is the expected number of PBHs within radius $r$,
\begin{equation}
N_{\rm pbh}(<r) \sim \frac{4\pi r^3}{3}\frac{\rho_{\rm pbh}(r)}{m_{\rm pbh}},
\label{eq:NPBH}
\end{equation}
which should satisfy $N_{\rm pbh}\gg 1$ near the characteristic feeding radius (often comparable to
the BH influence radius). If $N_{\rm pbh}\lesssim \mathcal{O}(1)$, the evolution is better described
as rare strong encounters rather than smooth diffusion; in this paper we restrict attention to the
diffusive regime.

\subsection{Collisionless refilling by triaxiality and chaotic centrophilic orbits}
\label{subsec:triaxial}

Spherical symmetry is not necessarily a robust idealization, especially at high redshift where halos are assembled
through mergers and are generically triaxial. In a non-spherical potential, angular momentum is no
longer an integral of motion, and collisionless torques can drive orbits to very small pericenters even
without collisionless relaxation. This produces a population of \emph{centrophilic} (often chaotic) orbits,
including pyramidal and related box-like orbits, which can be torqued to arbitrarily low angular momentum
in the smooth potential of the nucleus and thereby enter the loss cone
\cite{merritt2004,merritt2011,gualandris2017}.

Because the detailed chaotic dynamics depends on the degree of triaxiality, the central mass
concentration, and the radial range of interest, we adopt a minimal phenomenological description that
can be incorporated into Eq.~\eqref{eq:dMdot_interp}. We introduce an effective \emph{centrophilic fraction}
$f_{\rm ch}(\epsilon,t)$, defined as the fraction of phase space at energy $\epsilon$ occupied by
orbits that are efficiently  driven across the loss-cone boundary on a timescale of
order the orbital period. For these orbits the loss cone is approximately \emph{full}, implying
\begin{equation}
\left.\frac{f_{\rm lc}}{f_0}\right|_{\rm ch}\simeq 1
\qquad\Rightarrow\qquad
\left.\frac{d\dot M_{\rm BH}}{d\epsilon}\right|_{\rm ch}
\simeq
\frac{dM}{d\epsilon}\,\frac{R_{\rm lc}}{P}.
\label{eq:ch_fullLC}
\end{equation}
For the remaining fraction $(1-f_{\rm ch})$ of phase space, loss-cone feeding proceeds through
collisionless diffusion (stars and/or PBHs) with refilling parameter $q_{\rm diff}$.

This motivates the following composite capture spectrum:
\begin{equation}
\frac{d\dot M_{\rm BH}}{d\epsilon}
\simeq
\frac{dM}{d\epsilon}\,\frac{R_{\rm lc}}{P}
\left[
f_{\rm ch}(\epsilon,t)
+
\left(1-f_{\rm ch}(\epsilon,t)\right)
\frac{q_{\rm diff}(\epsilon,t)}{\ln\!\left[1/R_{\rm lc}(\epsilon,t)\right]+q_{\rm diff}(\epsilon,t)}
\right],
\label{eq:triax_mix}
\end{equation}
where
\begin{equation}
q_{\rm diff}(\epsilon,t)\equiv \frac{P(\epsilon,t)\,\mu_{\rm diff}(\epsilon,t)}{R_{\rm lc}(\epsilon,t)}.
\label{eq:qdiff}
\end{equation}

The parameter $f_{\rm ch}$ encapsulates the uncertain strength of collisionless torques. In a purely
triaxial halo one might expect $f_{\rm ch}$ to be nonzero over a broad energy range, while the growth
of a central BH can make the inner potential more nearly Keplerian and suppress centrophilic
trajectories within (or near) the BH sphere of influence. In practice we explore several simple
models, including:
\begin{align}
f_{\rm ch}(\epsilon,t) &= f_{\rm ch,0} \quad \text{(constant)}, \label{eq:fch_const}\\[3pt]
f_{\rm ch}(\epsilon,t) &= f_{\rm ch,0}\,\Theta\!\left[r_{\rm char}(\epsilon)-\eta\,r_{\rm inf}(t)\right],
\label{eq:fch_step}
\end{align}
where $\Theta$ is the Heaviside function, $r_{\rm char}(\epsilon)$ is the characteristic orbital
radius, $r_{\rm inf}(t)\equiv GM_{\rm BH}/\sigma^2$ is the BH influence radius~\cite{davis2014}, and $\eta\sim\mathcal{O}(1)$
controls where triaxiality is assumed to be suppressed.

\subsection{Combined refilling prescription}
\label{subsec:mu_total}

We decompose the collisionless (diffusive) refilling rate into a sum of independent contributions from stars and PBHs to the
diffusion coefficient,
\begin{equation}
\mu_{\rm diff}(\epsilon,t) \equiv \mu_\star(\epsilon,t) + \mu_{\rm pbh}(\epsilon,t),
\label{eq:mu_diff_sum}
\end{equation}
where $\mu_\star$ corresponds to stellar relaxation (when a stellar component is present) and
$\mu_{\rm pbh}$ corresponds to PBH-driven relaxation in the mixed DM scenario. Operationally we
compute each contribution from Eq.~\eqref{eq:trel_from_scattering} via the closure \eqref{eq:D_closure}
(or orbit averaging when required). The collisionless refilling from triaxiality enters separately
through $f_{\rm ch}$ as in Eq.~\eqref{eq:triax_mix}.

Equation~\eqref{eq:triax_mix} has two useful limiting behaviors. If $f_{\rm ch}\rightarrow 0$ it
reduces to the standard diffusion-limited loss-cone expression \eqref{eq:dMdot_interp}. If
$f_{\rm ch}\rightarrow 1$, the capture spectrum approaches the full-loss-cone (geometric) limit even
when the collisionless relaxation is inefficient. This parameterization therefore allows us to quantify the
extent to which collisionless non-sphericity can enhance early DM capture relative to spherical
models.

\section{Self-Consistent Time Evolution and Numerical Method}
\label{sec:numerics}

In this section we describe how we evolve the coupled BH--halo system when DM capture is not assumed
to be a small perturbation. The essential idea is to evolve the DM energy distribution under a
loss-cone sink term while updating the halo potential self-consistently, thereby capturing both the
depletion of tightly bound phase space and the deepening of the potential as $M_{\rm BH}$ grows.

\subsection{Quasi-static sequence of equilibria}
\label{subsec:quasistatic}

We assume that the system evolves through a sequence of near-equilibrium states such that the
dynamical time $t_{\rm dyn}(r)$ is short compared to the timescale on which $M_{\rm BH}$ and the halo
DF change appreciably. A useful diagnostic is the local growth time
\begin{equation}
t_{\rm grow}(t)\equiv \frac{M_{\rm BH}(t)}{\dot M_{\rm BH}(t)}.
\end{equation}
Our quasi-static approximation requires $t_{\rm grow}\gg P(\epsilon)$ for energies that dominate the
capture integral. When this condition holds, the halo rapidly phase-mixes in angle variables while
the DF in integrals of motion changes slowly, justifying the orbit-averaged loss-cone formalism.
Where $t_{\rm grow}\lesssim t_{\rm dyn}$, fully time-dependent methods (e.g. $N$-body or Vlasov--Poisson
solvers) would be required; we do not consider such regimes here.

\subsection{Evolving the energy distribution with a sink term}
\label{subsec:N_evolution}

Rather than evolve $f(\epsilon,t)$ directly, we evolve the mass per unit energy,
\begin{equation}
\mathcal{M}(\epsilon,t)\equiv \frac{dM}{d\epsilon}(\epsilon,t)=g(\epsilon,t)\,f(\epsilon,t),
\label{eq:N_def}
\end{equation}
which is strictly non-negative and integrates to the total DM mass represented by the DF,
\begin{equation}
M_{\rm DM}(t)=\int_0^\infty \mathcal{M}(\epsilon,t)\,d\epsilon.
\label{eq:MDM_from_N}
\end{equation}

At each time, the loss-cone theory yields a capture spectrum $S(\epsilon,t)\equiv d\dot M_{\rm BH}/d\epsilon$.
In the spherical case we use Eq.~\eqref{eq:dMdot_interp}. When including collisionless triaxial/chaotic
refilling we use the composite prescription Eq.~\eqref{eq:triax_mix} with a specified $f_{\rm ch}(\epsilon,t)$.
In all cases, the coupled evolution equations are
\begin{align}
\frac{\partial \mathcal{M}(\epsilon,t)}{\partial t} &= -S(\epsilon,t), \label{eq:N_sink}\\
\dot M_{\rm BH}(t) &= \int_0^\infty S(\epsilon,t)\,d\epsilon. \label{eq:Mdot_from_S}
\end{align}
Equations~\eqref{eq:N_sink}--\eqref{eq:Mdot_from_S} enforce mass conservation between the halo DF and
the BH sink (neglecting any additional sources not included in this paper, such as baryonic accretion).

\subsection{Self-consistent potential update}
\label{subsec:psi_update}

Given $\mathcal{M}(\epsilon,t)$ and $M_{\rm BH}(t)$, we determine the potential and density in a
self-consistent manner. For an isotropic DF, the density is computed from $f(\epsilon,t)=\mathcal{M}/g$ via
Eq.~\eqref{eq:rho_from_f}. The potential follows from spherical symmetry:
\begin{align}
M(<r,t) &= M_{\rm BH}(t) + 4\pi\int_0^r \rho(r',t)\,r'^2\,dr', \label{eq:Menc}\\
\psi(r,t) &= G \int_r^{r_{\max}} \frac{M(<r',t)}{r'^2}\,dr', \qquad \psi(r_{\max},t)=0,
\label{eq:psi_from_Menc}
\end{align}
which is equivalent to solving Poisson's equation with boundary condition $\psi(r_{\max})=0$.

The key complication is that $g(\epsilon,t)$ depends on $\psi(r,t)$ through Eq.~\eqref{eq:g_def}.
We therefore solve for $\psi(r,t)$ iteratively at each time step. Starting from an initial guess
$\psi^{(0)}(r)$ (typically the converged potential from the previous time step), we repeat:
\begin{enumerate}
\item Compute $g^{(k)}(\epsilon)$ from $\psi^{(k)}(r)$ using Eq.~\eqref{eq:g_def}.
\item Compute $f^{(k)}(\epsilon)=\mathcal{M}(\epsilon)/g^{(k)}(\epsilon)$.
\item Compute $\rho^{(k)}(r)$ from $f^{(k)}(\epsilon)$ and $\psi^{(k)}(r)$ via Eq.~\eqref{eq:rho_from_f}.
\item Compute an updated potential $\tilde\psi^{(k+1)}(r)$ from $\rho^{(k)}(r)$ and $M_{\rm BH}$ using
Eq.~\eqref{eq:psi_from_Menc}.
\item Under-relax to improve stability:
\begin{equation}
\psi^{(k+1)}(r) \leftarrow (1-\alpha)\,\psi^{(k)}(r) + \alpha\,\tilde\psi^{(k+1)}(r),
\qquad 0<\alpha\lesssim 1.
\label{eq:underrelax}
\end{equation}
\end{enumerate}
We iterate until the maximum fractional change in $\psi(r)$ falls below a tolerance (typically
$\lesssim 10^{-3}$--$10^{-4}$), and then proceed to compute $S(\epsilon,t)$.

This procedure self-consistently captures the backreaction of DM capture on the halo DF and density:
as $\mathcal{M}(\epsilon)$ is depleted at large binding energies, the central density tends to decrease.  This
feeds back by reducing the capture rate unless refilling from larger radii/energies is efficient.

\subsection{Orbital quantities for the loss cone}
\label{subsec:orbital_quantities}

The capture spectrum requires the period $P(\epsilon,t)$ and the circular-orbit angular momentum
$J_c(\epsilon,t)$, which appear in $R_{\rm lc}=J_{\rm lc}^2/J_c^2$ and in the refilling parameter
$q=P\mu/R_{\rm lc}$.

\paragraph{Loss-cone boundary.}
We adopt the Schwarzschild capture threshold $J_{\rm lc}\simeq 4GM_{\rm BH}/c$ (Eq.~\ref{eq:Jlc}) and
set
\begin{equation}
R_{\rm lc}(\epsilon,t)=\frac{J_{\rm lc}^2}{J_c^2(\epsilon,t)}.
\label{eq:Rlc_again}
\end{equation}

\paragraph{Circular orbits.}
Given $\psi(r,t)$, the circular speed is
\begin{equation}
v_c^2(r,t)=r\,\frac{d\psi}{dr},
\label{eq:vc2}
\end{equation}
and the circular-orbit energy at radius $r$ is
\begin{equation}
\epsilon_c(r,t)=\psi(r,t)-\frac{v_c^2(r,t)}{2}.
\label{eq:eps_circ}
\end{equation}
We construct a numerical inverse map $r_c(\epsilon,t)$ from $\epsilon_c(r,t)$ and then compute
\begin{equation}
J_c^2(\epsilon,t)=r_c^2(\epsilon,t)\,v_c^2[r_c(\epsilon,t),t].
\label{eq:Jc2}
\end{equation}

\paragraph{Orbital period.}
In the Keplerian regime, the period only depends upon $\epsilon$ and has the closed form of
Eq.~\eqref{eq:kepler_relations}. In the general self-gravitating case we define a characteristic
orbital period from the radial orbit ($J\rightarrow 0$),
\begin{equation}
P(\epsilon,t)\equiv 2\int_{0}^{r_{\max}(\epsilon,t)}\frac{dr}{\sqrt{2\,[\psi(r,t)-\epsilon]}}.
\label{eq:Prad}
\end{equation}
This choice is motivated by the fact that loss-cone feeding is dominated by low-$J$ trajectories and
ensures internal consistency with the same turning-point structure used in Eq.~\eqref{eq:g_def}.
Differences between alternative definitions of $P$ enter at order unity and do not qualitatively
change our conclusions.

\subsection{Refilling rate \texorpdfstring{$\mu(\epsilon,t)$}{mu(epsilon,t)} in time-dependent halos}
\label{subsec:mu_time}

For collisionless refilling channels, we compute a local relaxation time $t_{\rm rel}(r,t)$ using
Eq.~\eqref{eq:trel_from_scattering} with the appropriate perturber density $\rho_p(r,t)$ and mass $m_p$
(e.g. $\rho_p=\rho_{\rm pbh}$ and $m_p=m_{\rm pbh}$ for PBH-driven refilling). The corresponding
energy-dependent rate is then obtained via either the orbit average \eqref{eq:D_from_trel} or the
closure \eqref{eq:D_closure} with $r_{\rm char}=r_c(\epsilon,t)$.

When including collisionless refilling in non-spherical potentials, we use the composite capture
spectrum Eq.~\eqref{eq:triax_mix}. In practice this requires specifying a model for
$f_{\rm ch}(\epsilon,t)$ [e.g. Eq.~(\ref{eq:fch_const}) or (\ref{eq:fch_step})] and then evaluating
$q_{\rm diff}=P\mu_{\rm diff}/R_{\rm lc}$ for the remaining phase-space fraction
$(1-f_{\rm ch})$.

\subsection{Discretization and timestep control}
\label{subsec:discretization}

We discretize the radius and energy on logarithmic grids,
\begin{equation}
r_i \in [r_{\min},r_{\max}],\qquad \epsilon_j \in [\epsilon_{\min},\epsilon_{\max}],
\end{equation}
with quadrature weights chosen to accurately resolve the integrals in
Eqs.~\eqref{eq:rho_from_f}, \eqref{eq:g_def}, and \eqref{eq:Prad}. At each timestep $\Delta t$ we
compute $S_j=S(\epsilon_j,t)$ and update
\begin{equation}
\mathcal{M}_j(t+\Delta t)=\max\!\left[\mathcal{M}_j(t)-S_j\,\Delta t,\,0\right],
\qquad
M_{\rm BH}(t+\Delta t)=M_{\rm BH}(t)+\left(\sum_j S_j\,w_j\right)\Delta t,
\label{eq:discrete_update}
\end{equation}
where $w_j$ are the energy quadrature weights.

To ensure numerical stability and maintain the quasi-static assumption, we use an adaptive time step
chosen such that no energy bin is depleted by more than a fixed fraction per step:
\begin{equation}
\Delta t = \min\!\left[\Delta t_{\max},\ f_{\rm step}\,\min_j\left(\frac{\mathcal{M}_j}{S_j}\right)\right],
\qquad 0<f_{\rm step}\ll 1.
\label{eq:dt_control}
\end{equation}
We also enforce a minimum timestep floor to avoid pathological behavior when $S_j\rightarrow 0$.

\section{Results}\label{sec:results}

In this section we present the DM-driven contribution to early SMBH growth in the loss-cone framework
developed in Sections~\ref{sec:setup}--\ref{sec:numerics}. Our emphasis is on identifying the
conditions under which collisionless DM capture is non-negligible by $z\sim 10$, and on clarifying
the roles of (i) collisionless refilling by stars and/or PBHs as massive perturbers, (ii) collisionless
refilling by triaxial/chaotic orbits, and (iii) the effect of capture on the halo DF and potential.

In Section~\ref{subsec:pbh_refill} it was shown that in the diffusive regime
($N_{\rm pbh}\gg 1$ in the feeding region), the PBH-driven relaxation rate scales as
$\mu_{\rm pbh}\propto m_{\rm pbh}\rho_{\rm pbh}\propto f_{\rm pbh}m_{\rm pbh}\rho_{\rm DM}$ (Eq.~\ref{eq:scaling_pbh}).
Accordingly, the key control parameter is the \emph{granularity amplitude $\Xi$}:
\begin{equation}
\Xi \equiv f_{\rm pbh}\,m_{\rm pbh},
\label{eq:Xi_def}
\end{equation}
which enters $t_{\rm rel}$, and hence $q$, at fixed $\rho_{\rm DM}$ and $\sigma$.

\subsection{Fiducial model suite}\label{subsec:fiducial-suite}

Table~\ref{tab:fiducial_models} defines four fiducial cases designed to isolate the impact of
(i) the refilling agent (stars versus PBHs) and (ii) collisionless centrophilic orbit families in a
non-spherical potential, modeled here by the phenomenological parameter $f_{\rm ch}$ (Section~\ref{subsec:triaxial}).
All fiducial runs start from a $M_{{\rm BH},0}=10^{5}\,M_\odot$ seed in an NFW halo with
virial mass $M_{\rm vir}=10^8\,M_\odot$ and concentration $c=3$. The ``Sph--Star'' case adopts a less
compact configuration ($r_{1/2}=1~{\rm pc}$) with stellar two-body relaxation as the only refilling mechanism.
This case is
intended as a conservative stellar-refilling baseline rather than an extreme
compact-cluster model. If stars traced the DM profile,
\begin{equation}
    \rho_\star/\rho_{\rm DM}\simeq \epsilon_\star f_b/(1-f_b),
\end{equation}
where \(\rho_\star\) is the stellar density, \(f_b\) is the cosmic baryon fraction, and \(\epsilon_\star\)
is the integrated star-formation efficiency. For plausible high-redshift systems we expect
\(\epsilon_\star\ll1\), especially in low-mass halos~\cite{tacchella2018,sun2016}. A stellar component
that traces the DM halo therefore has \(\rho_\star/\rho_{\rm DM}\ll 1\), making stellar refilling negligible.
A compact nuclear stellar component can instead have \(\rho_\star\sim \rho_{\rm DM}\) at parsec radii.
Observed nuclear star clusters and young massive clusters generally have characteristic radii of order a
few parsecs. For example, late-type nuclear star clusters have a median effective radius $r_e\simeq 3.5$ pc,
with roughly half the sample between
$2.4$ and $5.0$ pc, while young massive clusters have effective radii
spanning $\sim0.5$--$10$ pc and peaking near $2$--$3$ pc
\cite{boker2004,ryon2017}. A stellar system compressed to
$r_{1/2}=0.1$ pc would be so dense that stellar collisions, core collapse,
binary heating, runaway mergers, and TDEs could dominate the evolution
\cite{portegies2010}. Such processes are outside
the collisionless DM loss-cone framework considered in this paper. Thus
$r_{1/2}=1$ pc is already compact but avoids conflating ordinary stellar
relaxation with a separate collisionless stellar-runaway channel.

The PBH-driven cases adopt a much more compact configuration ($r_{1/2}=0.1~{\rm pc}$) and include a PBH
subcomponent with $(f_{\rm PBH},m_{\rm PBH})=(0.1,1\,M_\odot)$, i.e. $\Xi=0.1\,M_\odot$.
The triaxial/chaotic refilling models considered include $f_{\rm ch}=0.5$ and $f_{\rm ch}=1$, where the latter
corresponds to a full-loss-cone (geometric) supply at fixed energy.

\vspace{6pt}
\noindent
\textit{Figure set and organization.}
In what follows we refer to (i) growth histories $M_{\rm BH}(t)$ and $\dot M_{\rm BH}(t)$
(Fig.~\ref{fig:growth_histories}), (ii) the impact of PBH granularity and the initial seed mass
$M_{{\rm BH},0}$ (Fig.~\ref{fig:param_scan}), and (iii) halo redistribution and depletion of the central DF
(Fig.~\ref{fig:backreaction}).

\begin{table*}[t]
\centering
\caption{Fiducial models}
\resizebox{\textwidth}{!}{\begin{tabular}{c c c c c c}
\hline
Model & $M_{{\rm BH},0}$ [$M_\odot$] & Halo $(M_{\rm vir}[M_\odot],r_{1/2}[\rm pc],c)$ & Refilling & $(f_{\rm pbh},m_{\rm pbh}[M_\odot])$ & $f_{\rm ch}$ model \\
\hline
Sph--Star  & $10^5$ & $(10^8,1,3)$ & stars only & -- & $0$ \\
Sph--PBH   & $10^5$ & $(10^8,0.1,3)$ & PBH only   & $(0.1,1)$ & $0$ (spherical potential) \\
Tri--PBH  & $10^5$ & $(10^8,0.1,3)$ & PBH only & $(0.1,1)$ & $0.5$ \\
Tri--PBH   & $10^5$ & $(10^8,0.1,3)$ & PBH only   & $(0.1,1)$ & $1$ (full loss cone)  \\
\hline
\end{tabular}}
\label{tab:fiducial_models}
\end{table*}

\subsection{Representative SMBH growth histories}\label{subsec:fiducial-histories}

\begin{figure*}[t]
\centering
\begin{subfigure}{.5\textwidth}
			\centering
			\includegraphics[width=1\linewidth]{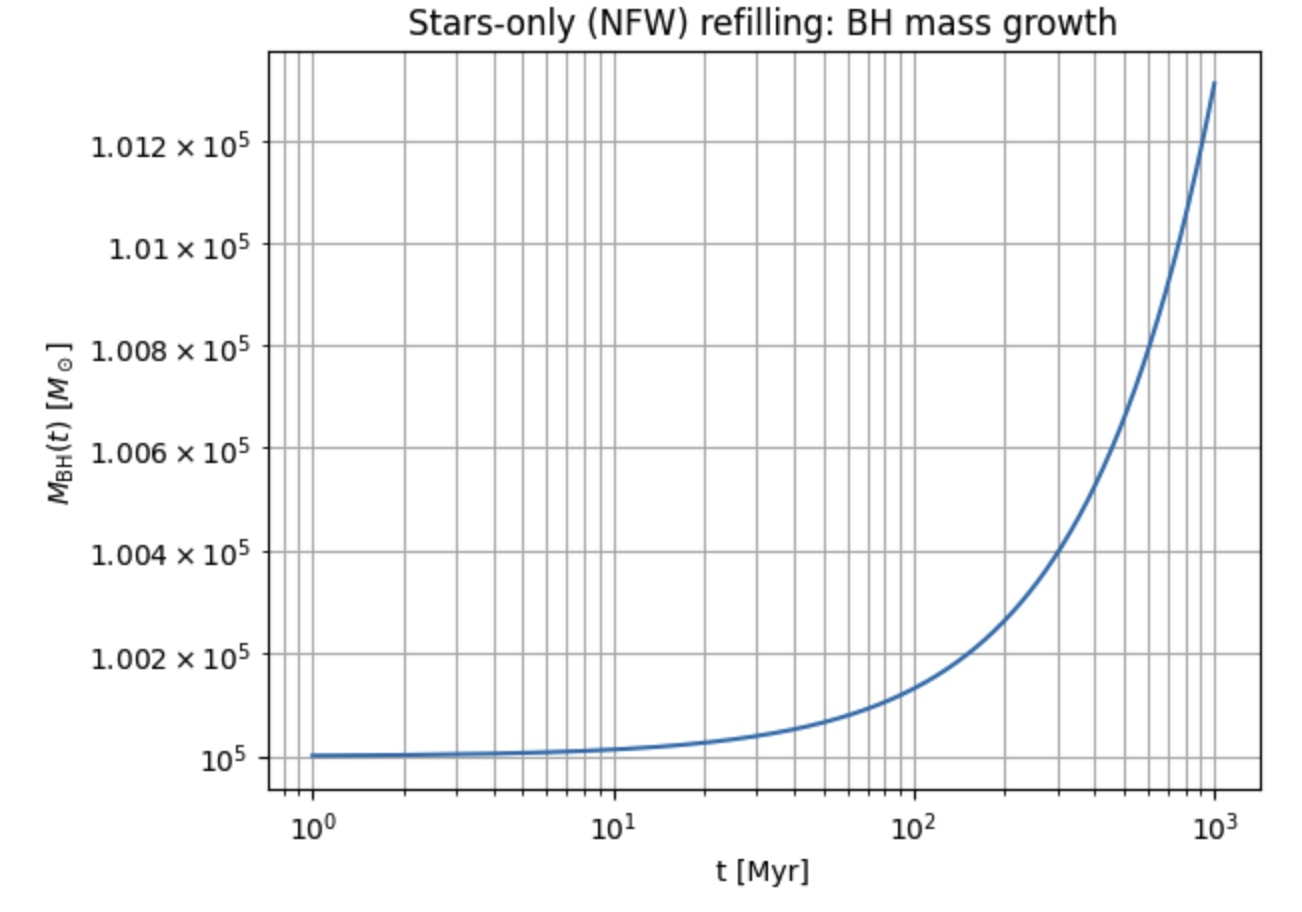}
		\end{subfigure}%
	\begin{subfigure}{.5\textwidth}
			\centering
			\includegraphics[width=1\linewidth]{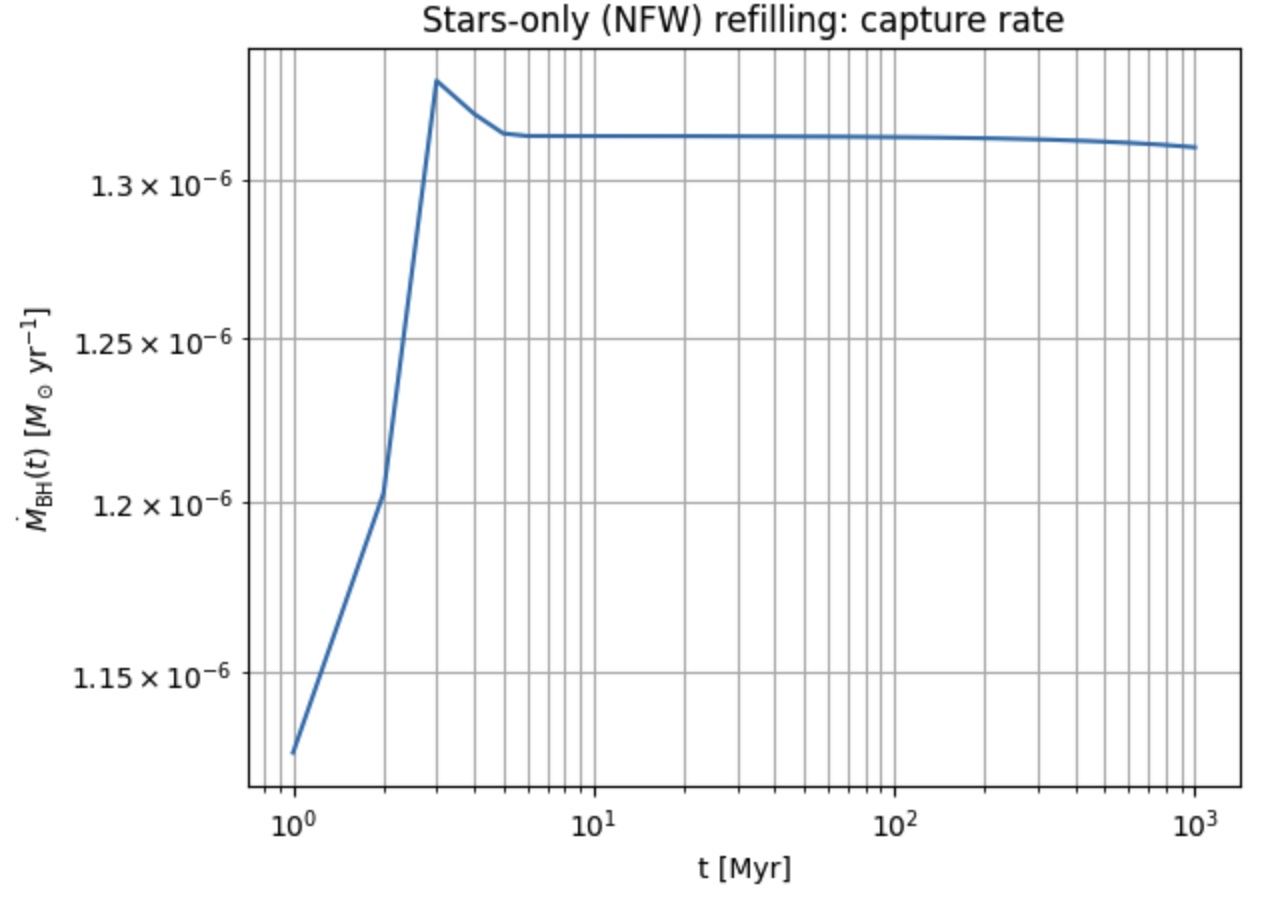}
		\end{subfigure}
            \begin{subfigure}{.5\textwidth}
			\centering
			\includegraphics[width=1\linewidth]{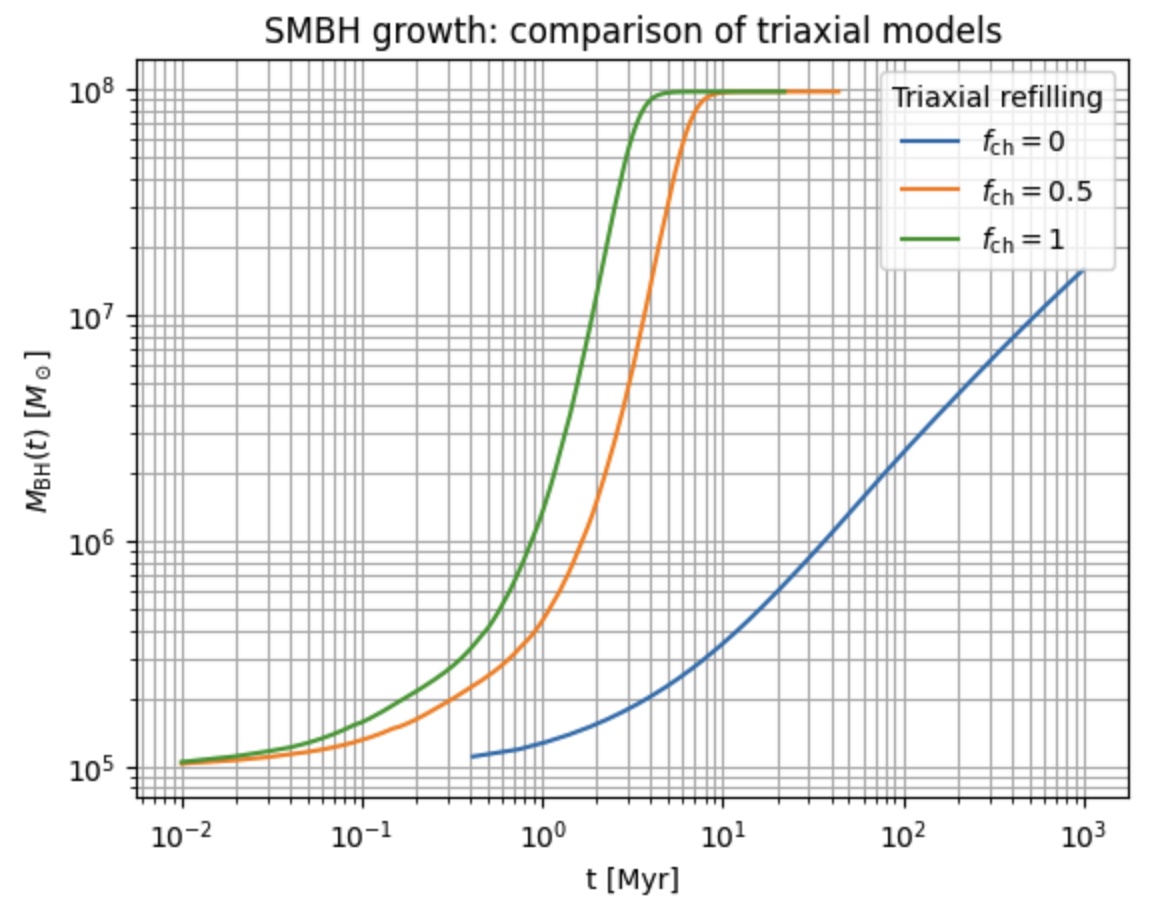}
		\end{subfigure}%
		\begin{subfigure}{.5\textwidth}
			\centering
			\includegraphics[width=1\linewidth]{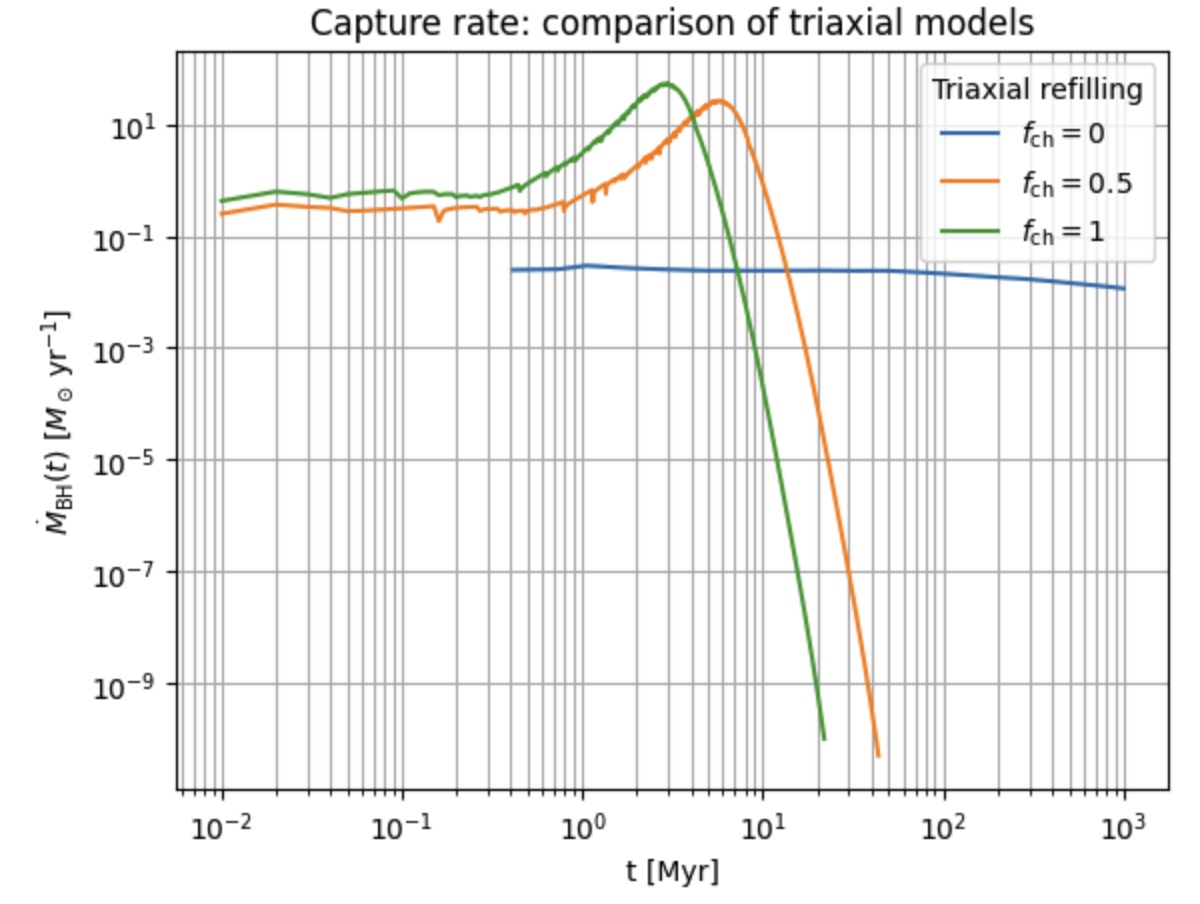}
		\end{subfigure}
        \caption{
Representative SMBH growth histories $M_{\rm BH}(t)$ (left) and $\dot M_{\rm BH}(t)$ (right) for the
fiducial seed and halo in Table~\ref{tab:fiducial_models}. Curves compare stars-only refilling (top), PBH-driven
refilling, and triaxial
models with nonzero $f_{\rm ch}$ (bottom). The model with $f_{\rm ch}=0$ corresponds to a spherically symmetric potential and the model with $f_{\rm ch}=1$ corresponds to full-loss-cone (geometric) supply.}
\label{fig:growth_histories}
\end{figure*}

Figure~\ref{fig:growth_histories} summarizes the characteristic time dependence of both the SMBH mass
$M_{\rm BH}(t)$ and the capture rate $\dot M_{\rm BH}(t)$ in the fiducial runs.

\paragraph{Stars-only refilling is inefficient.}
The top row of Figure~\ref{fig:growth_histories} shows the ``Sph--Star'' model.
The capture rate remains at $\dot M_{\rm BH} \sim 10^{-6}\,M_\odot\,{\rm yr}^{-1}$ over the plotted interval,
leading to only percent-level growth of a $10^5\,M_\odot$ seed even over $\gtrsim{\rm Gyr}$ timescales.
This establishes a useful baseline: in the absence of a very dense granular/heavy scattering population (or strong
non-spherical orbital transport), collisionless refilling by ordinary stars does not drive appreciable
DM capture for the adopted halo parameters.

\paragraph{PBH-driven refilling produces substantial growth in spherical symmetry.}
The bottom row of Figure~\ref{fig:growth_histories} compares PBH-driven models with different $f_{\rm ch}$.
In the spherical diffusion-limited case ($f_{\rm ch}=0$), the SMBH grows gradually over $\sim10^2$--$10^3$ Myr,
with a quasi-steady capture rate of order $\dot M_{\rm BH} \sim 10^{-2}$--$10^{-1}\,M_\odot\,{\rm yr}^{-1}$.
For the compact fiducial halo ($r_{1/2}=0.1~{\rm pc}$) and $\Xi=0.1\,M_\odot$, this would result in
one to several orders of magnitude of DM-driven growth by high redshift (consistent with the scan in
Figure~\ref{fig:param_scan}; see below).

\paragraph{Triaxial/chaotic refilling yields bursty, supply-limited growth.}
Allowing a nonzero fraction of centrophilic orbits ($f_{\rm ch}>0$) qualitatively changes the evolution.
For $f_{\rm ch}=0.5$ and $f_{\rm ch}=1$, the SMBH experiences an early-time burst in $\dot M_{\rm BH}$ that can
reach $\sim 1$--$10^{1}\,M_\odot\,{\rm yr}^{-1}$, rapidly increasing the BH mass to
$M_{\rm BH} \sim 10^{8}\,M_\odot$ in a few Myr.

The fiducial histories in Figure~\ref{fig:growth_histories} adopt
$M_{\rm BH,0}=10^{5}\,M_\odot$, which should be interpreted as an
optimistic heavy-seed benchmark rather than a generic initial condition.
Such a seed mass is motivated by rare direct-collapse or supermassive-star
formation scenarios, but more standard light-seed channels, such as Pop~III
remnants, would instead produce seeds with
$M_{\rm BH,0}\sim 10^2$--$10^3\,M_\odot$. Since the efficiency of
loss-cone DM capture depends on both the initial BH mass and the available
phase-space reservoir, we test how strongly our conclusions depend on the assumed seed mass. In the next
subsection we perform a sensitivity scan over $M_{\rm BH,0}$, and we focus on the final
DM-driven BH mass at $z=10$,
\begin{equation}
M_{\rm BH,10}\equiv M_{\rm BH} (z=10).
\end{equation}

\subsection{Parameter dependence in spherical models}
\label{subsec:scan}
\begin{figure*}[t]
\centering
\includegraphics[width=\textwidth]{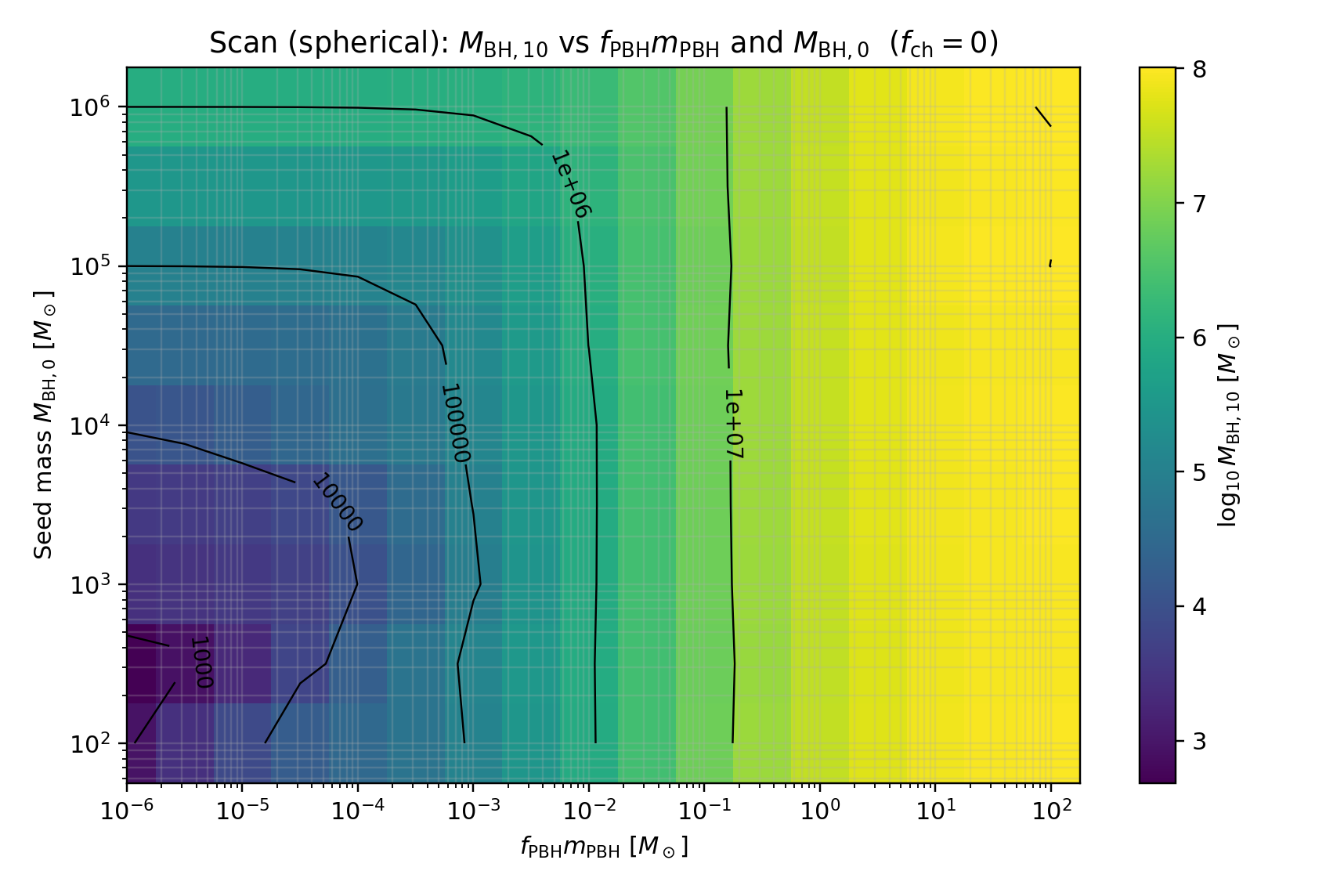}
\caption{
Parameter scan showing the final DM-driven BH mass at \(z=10\),
\(M_{\rm BH,10}\), as a function of PBH granularity
\(\Xi=f_{\rm PBH}m_{\rm PBH}\) and initial seed mass \(M_{\rm BH,0}\) at \(z=30\),
for spherical models with \(f_{\rm ch}=0\).  Contours indicate fixed
values of \(M_{\rm BH,10}\).}
\label{fig:param_scan}
\end{figure*}

Figure~\ref{fig:param_scan} presents a two-dimensional parameter scan for spherical models with $f_{\rm ch}=0$. The color coding  shows the final BH mass at \(z=10\),
\(M_{\rm BH,10}\), as a function of the PBH granularity parameter
\(\Xi\) and the initial seed mass \(M_{\rm BH,0}\) at $z=30$.  This scan complements the
fiducial growth histories by allowing both light-seed and heavy-seed
initial conditions to be compared within the same loss-cone framework. Growth increases strongly with granularity.
At fixed $M_{{\rm BH},0}$, increasing $\Xi$ produces a monotonic increase in \(M_{\rm BH,10}\), reflecting the
shortening of the relaxation time and the corresponding increase in the loss-cone flux. More interestingly, the final mass \(M_{\rm BH,10}\) becomes weakly dependent on
the initial seed mass over a broad range of parameter space, even before the system
reaches the full-loss-cone limit. This behavior follows from the diffusion-limited form of the loss-cone flux. When \(q\ll \ln(1/R_{\rm lc})\), the explicit BH-mass factor
\(R_{\rm lc}/P\) in the spherical capture spectrum, Equation~(\ref{eq:dMdot_interp}), cancels and the flux is set mainly by the orbit-averaged
diffusion rate \(\mu\) and the available phase-space mass \(\mathcal{M}(\epsilon,t)\). Consequently, over a fixed time interval the BH gains an approximately
seed-independent mass \(M_{\rm acc}^{\rm diff}\). If
\(M_{\rm acc}^{\rm diff}\gg M_{\rm BH,0}\), then
\(M_{\rm BH,10}\simeq M_{\rm acc}^{\rm diff}\), producing nearly vertical
contours in the \(M_{\rm BH,10}\) scan even though the loss cone is not yet full.

The density of states \(g(\epsilon)\) is not generally independent of \(M_{\rm BH}\). In fact,
Eq.~(\ref{eq:g_simplified}) shows that in a Keplerian  potential
\(g(\epsilon)\propto M_{\rm BH}^3\). For a fixed physical density profile, however, the distribution
function obtained by Eddington inversion changes with \(M_{\rm BH}\) in such a way that the mass in orbits
of a given physical scale is nearly unchanged. For example, for a cusp \(\rho\propto r^{-\gamma}\) in a
Keplerian  potential, one finds \(g f\,d\epsilon \propto a^{2-\gamma}da\), independent of \(M_{\rm BH}\),
where \(a=GM_{\rm BH}/(2\epsilon)\) is the semi-major axis.

Therefore, Fig.~\ref{fig:param_scan} provides a convenient ``feasibility map'':
regions at small $\Xi$ correspond to negligible DM-driven growth (diffusion too slow),
while regions at larger $\Xi\sim 0.01$--$1\,M_\odot$ allow SMBHs with masses of $\sim 10^{6}$--$10^{8}\,M_\odot$ to form by $z\sim 10$ even with light seeds and without invoking
triaxial/chaotic orbit families.
(Including $f_{\rm ch}>0$ would shift the effective growth contours toward smaller $\Xi$ by increasing the
effective supply to the loss cone at fixed energy.)

\subsection{Fixed background versus self-consistent evolution}\label{subsec:fixed-vs-selfconsistent}

\begin{figure*}[t]
\centering
\begin{subfigure}{.5\textwidth}
			\centering
			\includegraphics[width=1\linewidth]{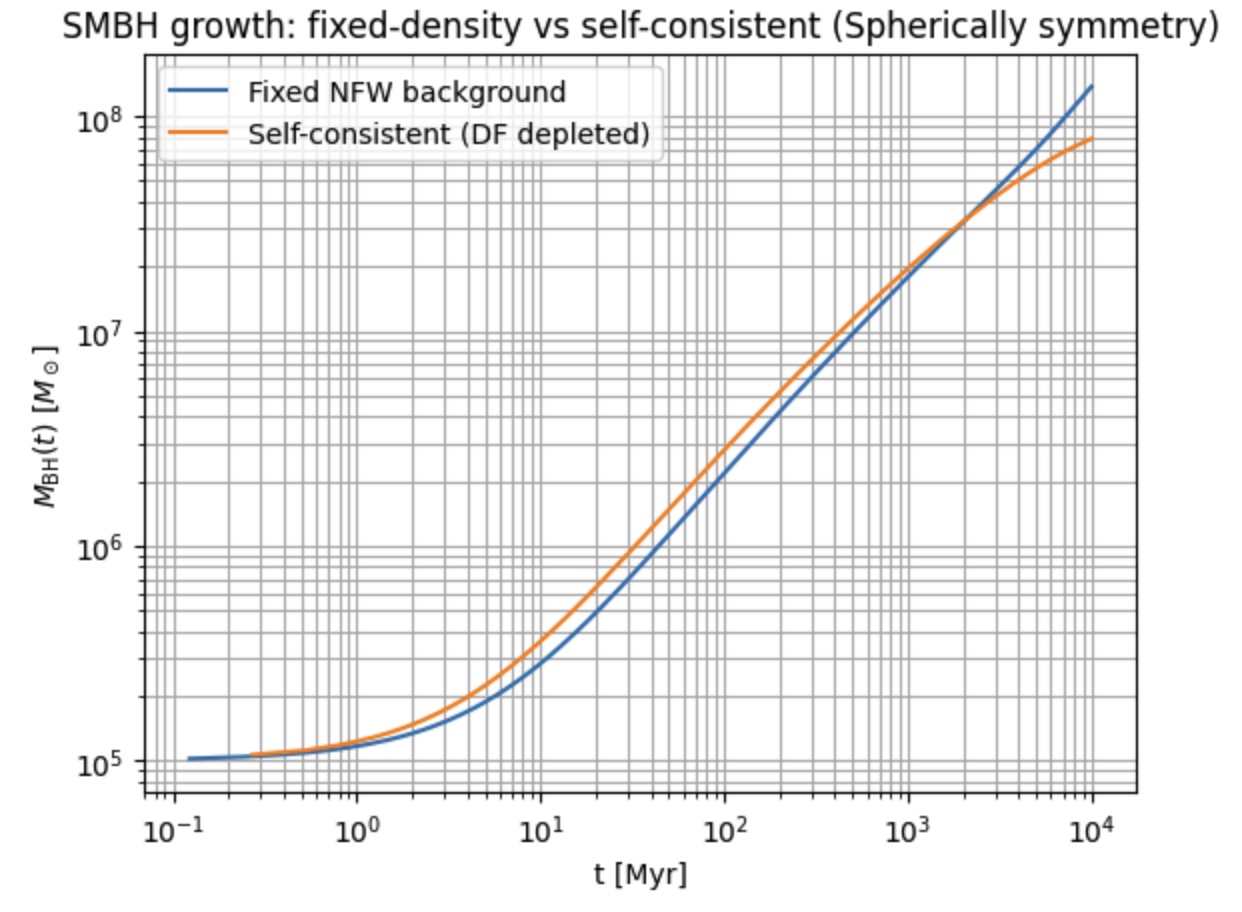}
		\end{subfigure}%
	\begin{subfigure}{.5\textwidth}
			\centering
			\includegraphics[width=1\linewidth]{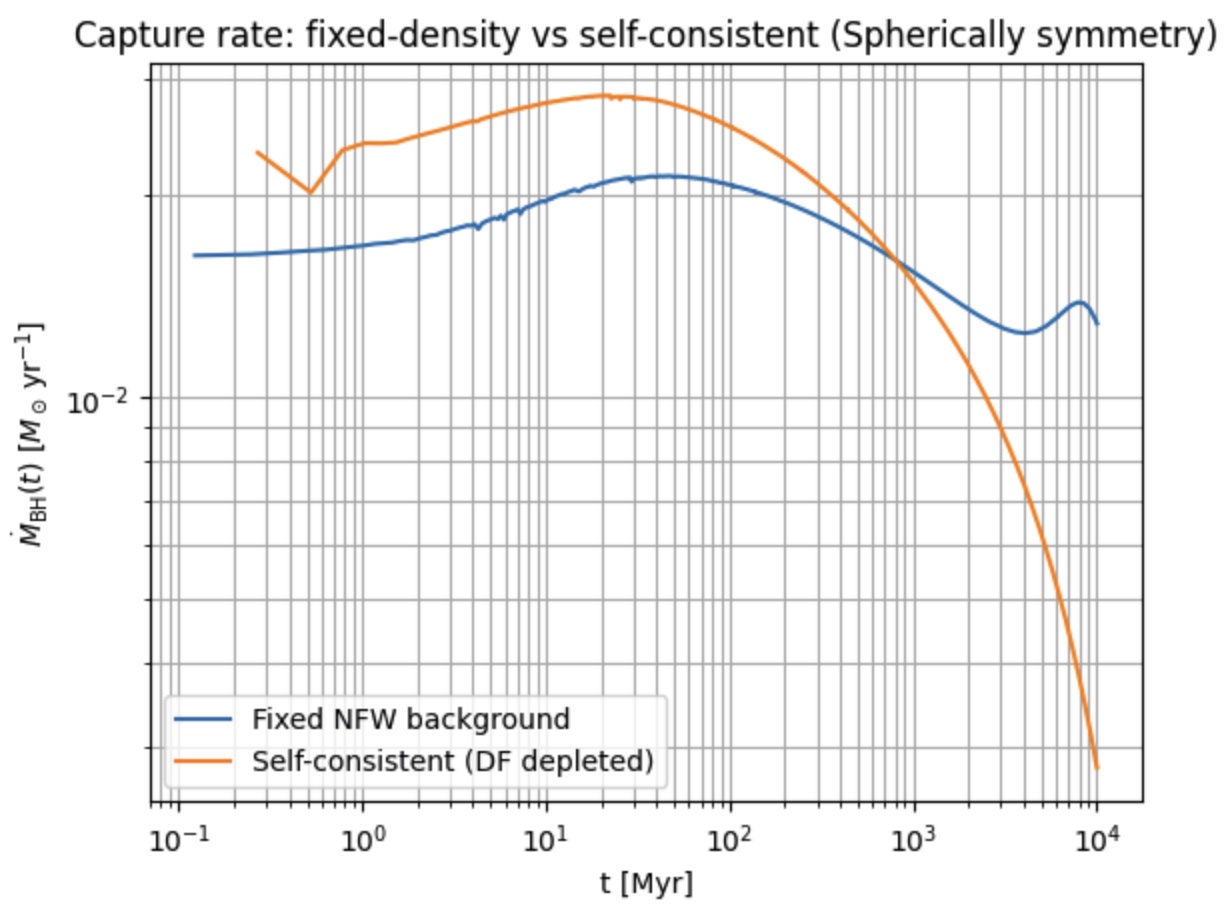}
		\end{subfigure}
    \begin{subfigure}{.5\textwidth}
			\centering
			\includegraphics[width=1\linewidth]{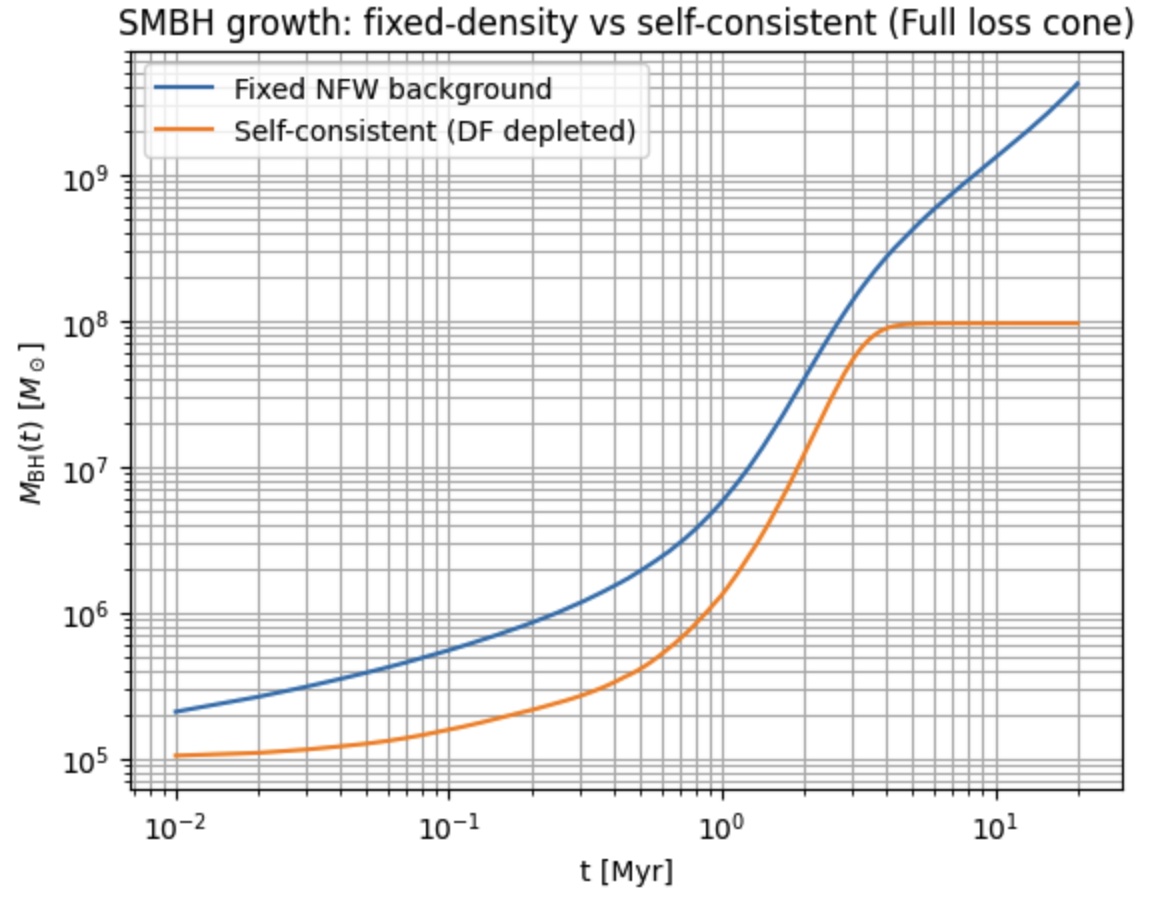}
		\end{subfigure}%
	\begin{subfigure}{.5\textwidth}
			\centering
			\includegraphics[width=1\linewidth]{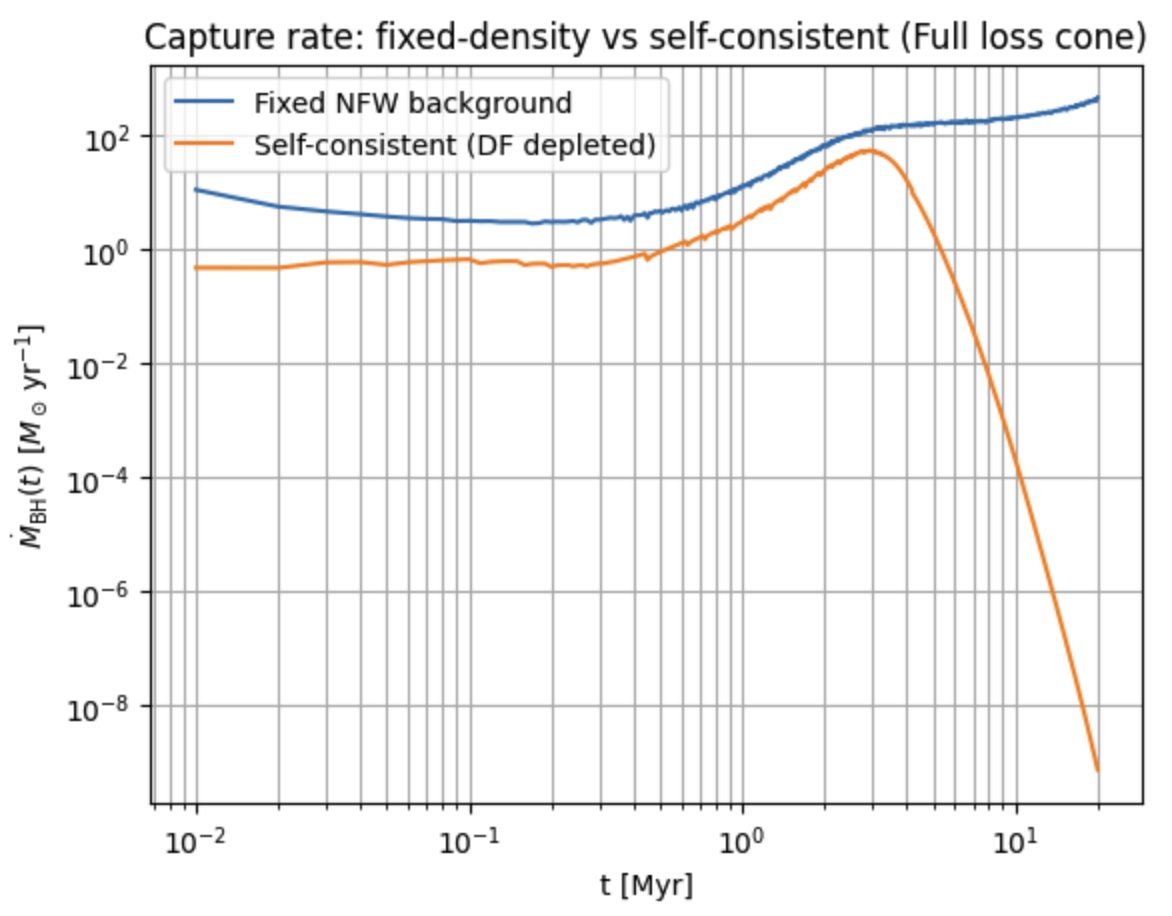}
		\end{subfigure}
\caption{
Comparison of fixed-density (NFW) background and self-consistent time evolution models with identical refilling prescriptions for spherical (top) and full-loss-cone (bottom) cases.}
\label{fig:backreaction}
\end{figure*}

Figure~\ref{fig:backreaction} isolates the impact of halo backreaction and phase-space depletion by
comparing two numerical treatments under otherwise identical refilling prescriptions:
(i) a ``fixed-density'' calculation in which the NFW background is held fixed (effectively an infinite
reservoir) and (ii) a self-consistent calculation in which the distribution function is depleted by capture
and the potential is updated accordingly.

\paragraph{Diffusion-limited (spherical) evolution: modest differences at early times, larger at late times.}
In the top row of Figure~\ref{fig:backreaction} (spherical symmetry), the two approaches agree
reasonably well at early times when the captured mass is a small perturbation to the halo.
At later times the curves diverge: the self-consistent capture rate declines as the low-$J$ reservoir is
depleted, while the fixed-background model maintains a comparatively high $\dot M_{\rm BH}$ because it
implicitly replenishes the depleted phase space.
The corresponding BH mass in the fixed-background model is therefore systematically overestimated at late
times.  (A small early-time offset, visible in $\dot M_{\rm BH}$ at $\lesssim{\rm Myr}$, is expected in practice
because the discretized DF--potential pair in the self-consistent iteration can differ slightly from the
analytic input NFW profile at very small radii; this transient does not affect the qualitative conclusions.)

\paragraph{Full loss cone: fixed-background runs strongly overestimate SMBH growth.}
The bottom row of Figure~\ref{fig:backreaction} demonstrates that the fixed-background
approximation becomes unreliable in the full-loss-cone limit.
When the loss cone is forced to remain full at all energies, the fixed-background model predicts a runaway
increase of $\dot M_{\rm BH}$ and correspondingly rapid growth to $M_{\rm BH}\gtrsim 10^9\,M_\odot$ on
timescales of a few--$10$ Myr.
In contrast, the self-consistent model exhibits a short burst followed by a sharp decline in $\dot M_{\rm BH}$,
and the BH mass saturates at $M_{\rm BH}\sim 10^8\,M_\odot$ once the accessible DF reservoir is consumed.
Thus, in the most optimistic refilling scenarios (large $f_{\rm ch}$), self-consistent depletion is the
dominant regulator of SMBH growth: the limiting factor is not the instantaneous angular-momentum transport
rate, but the finite amount of mass available in the portion of phase space that can be driven into the loss cone.

\subsection{A TNG50-calibrated benchmark}
\label{subsec:tng50_benchmark}

The compact fiducial halos in Table~\ref{tab:fiducial_models} are designed to explore the maximum possible
impact of DM loss-cone capture in rare, dense environments.  To assess how these idealized models compare
with a more cosmological density profile, we also apply the same loss-cone framework to a DM profile
extracted from the inner $\sim 200$ pc of a TNG50 galaxy progenitor at $z=20$, whose descendant hosts a
$6\times 10^8\,M_\odot$ SMBH at $z=0$.  The extracted profile is well fit by an NFW form with~\cite{Imai2026}
\begin{equation}
\rho_s \simeq 2.49\pm1.05\,M_\odot\,{\rm pc}^{-3},
\qquad
r_s\simeq 210.63\pm 38.62\,{\rm pc}.
\end{equation}

We find that this TNG50-calibrated NFW profile leads to negligible DM-driven growth of a
$10^5\,M_\odot$ seed by $z\simeq 7$. Even in the optimistic full-loss-cone limit, the SMBH mass increases
only by a tiny fraction of its initial value. This follows directly from the scaling
\begin{equation}
\dot M_{\rm BH} \propto \rho_{\rm DM}\,M_{\rm BH}^2/\tilde v ,
\end{equation}
which implies that for $\tilde v\sim100\,{\rm km}\,{\rm s}^{-1}$, one would need
$\rho_{\rm DM}\sim 10^8\,M_\odot\,{\rm pc}^{-3}$ to grow a seed from $10^5\,M_\odot$ to
$\sim10^7\,M_\odot$ between $z\sim 20$ and $z\sim 7$. Even extrapolating the NFW profile inward, the
fitted TNG50 density at parsec scales is only of order
\begin{equation}
\rho(r)\sim \rho_s \frac{r_s}{r}\sim (530\pm 360)\left(\frac{1\,\rm pc}{r}\right)\,M_\odot\,{\rm pc}^{-3},
\end{equation}
which is many orders of magnitude below the densities required for rapid collisionless capture of DM onto a
$10^5\,M_\odot$-scale seed.

This result provides an important realism check: DM loss-cone capture is not generically efficient in
ordinary NFW-like cosmological halos.  The large growth found in our compact fiducial models should
therefore be interpreted as an upper-envelope or special-environment scenario, requiring a much denser
nuclear DM configuration than is present in this TNG50 benchmark.

We caution, however, that the TNG50 profile is not a direct measurement of the sub-pc DM density relevant
for capture.  The fit is constrained mainly at radii of tens to hundreds of parsecs, while the capture
process is sensitive to the unresolved central phase space. For example, if DM is clustered with the
Pseudo-Jaffe profile~\cite{powell2025,yu2026}, the TNG50-fitted density at small $r$ would be~\cite{Imai2026}
\begin{equation}
    \rho(r)\sim(2.6\pm 1.4)\times 10^4 \left(\frac{1\, {\rm pc}}{r}\right)^2\,M_\odot\,{\rm pc}^{-3},
\end{equation}
which yields the required densities for $r\lesssim 0.01\, \rm pc$. Thus the TNG50 benchmark does not rule out
growth in environments with adiabatic contraction, DM spikes, compact PBH clusters, or other processes that
increase the central phase-space density (cf.~Ref.~\cite{Imai2026}).  Rather, it shows that a simple extrapolation of a resolved
NFW-like cosmological profile is insufficient.

\section{Conclusions}
\label{sec:conclusions}

We have a developed new framework to describe the loss-cone dynamics of the  gravitational capture of collisionless DM onto a central seed black hole. We have then analyzed the impact of this framework on models for the contribution of DM capture toward the early formation of SMBHs.
In this framework, the capture rate is controlled by (i) the angular-momentum boundary of the relativistic
loss cone, (ii) the orbital properties of the surrounding phase-space distribution, and (iii) the
mechanism that refills low-angular-momentum states. We included stellar two-body relaxation as a baseline for collisional
loss-cone refilling. We found that DM-driven growth in the stellar-refilling benchmark is limited, but this applies only
to the adopted stellar and halo parameters and should not be interpreted
as a general exclusion of efficient refilling by stars or stellar remnants
in other nuclear environments.
Motivated by mixed-DM scenarios, we emphasized refilling by a granular population of PBHs, and also explored an additional collisionless refilling channel in non-spherical potentials,
parameterized phenomenologically by a ``chaotic/triaxial'' fraction $f_{\rm ch}$. Stellar remnants can contribute within the same gravitational-scattering
framework, although their populations and dynamical evolution are not
modeled separately in the present study.

A wide range of astrophysical and cosmological probes constrain the allowed PBH fraction $f_{\rm PBH}(m)$,
including microlensing surveys, dynamical heating/evaporation of stellar systems, disruption of wide
binaries, CMB anisotropy constraints from PBH accretion, and limits from gravitational-wave merger rates~\cite{carr2021}.
Under standard assumptions, these constraints typically imply that \emph{stellar-mass} PBHs constitute at
most a subdominant fraction of the DM, i.e. $f_{\rm PBH}\ll 1$ for $m_{\rm PBH}\sim M_\odot$.
Our analysis nevertheless shows that PBHs could
act primarily as \emph{granular scatterers} that drive loss-cone refilling even when they constitute only a
subdominant component of the overall DM density.
This motivates expressing results in terms of a granularity parameted $\Xi$, which can be mapped onto observationally allowed
regions of $(m_{\rm PBH},f_{\rm PBH})$ once a specific constraint compilation is adopted.

Taken together, these results demonstrate the new result that DM loss-cone capture can, under special circumstances,
provide a rapid, non-radiative contribution to early SMBH growth.
The mechanism is most efficient in compact, high-density environments and in the presence of strong
angular-momentum transport, either collisionless (PBH granularity) or collisionless (triaxial/chaotic
orbits).  

Several important extensions are left to future work.
First, embedding the model into a cosmological halo assembly history will determine how frequently the
required compact nuclear configurations arise and persist.
Second, coupling DM capture to baryonic processes (gas inflow, star formation, and stellar tidal
disruption) will enable a unified picture in which DM capture acts as an early-time accelerator that may
reduce the burden on later luminous accretion.
Third, replacing the phenomenological $f_{\rm ch}$ parameterization with orbit-based transport in explicit
non-spherical potentials will sharpen predictions for the most optimistic refilling scenarios.
Finally, translating the $\Xi$-based scan into explicitly allowed regions using up-to-date PBH constraints
will clarify whether the PBH-driven refilling channel can operate at the level required to impact the
observed population of high-redshift SMBHs.

In summary, we have shown on the basis of a new application of loss-cone dynamics that DM-driven loss-cone capture is not expected to dominate SMBH growth in generic environments.  However, 
 it could  be dynamically important in dense systems with efficient refilling, and it provides a
useful theoretical limiting case for assessing the maximum possible ``dark'' contribution to early SMBH
assembly.

\acknowledgments

Work supported by the U.S.  Department of Energy under Nuclear Theory Grant DE-FG02-95-ER40934.

\paragraph{Author contribution.} All authors contributed equally to this work.


\bibliographystyle{plain}
\bibliography{references}

% Please avoid comments such as "For a review'', "For some examples",
% "and references therein" or move them in the text. In general,
% please leave only references in the bibliography and move all
% accessory text in footnotes.

% Also, please have only one work for each \bibitem.

\end{document}